\documentclass[superscriptaddress,groupedaddress,nofootnoteinbib,11pt]{article}
\pdfoutput=1

\usepackage[T1]{fontenc}
\usepackage[utf8]{inputenc}
\usepackage{upgreek}
\usepackage{amsmath,amssymb,mathtools}
\usepackage{mathrsfs,amsfonts,bm}
\usepackage{slashed}
\usepackage{tensor}
\usepackage{physics}
\usepackage{dsfont}
\usepackage{hepunits}
\usepackage{comment,cancel}

\usepackage{graphicx,epsfig,psfrag}
\usepackage{subfig}
\usepackage{tikz}
\usepackage{tikz-feynman}
\usetikzlibrary{tikzmark}

\PassOptionsToPackage{table,xcdraw,dvipsnames,usenames}{xcolor}
\usepackage{jheppub}
\usepackage{colortbl}
\usepackage{lmodern}
\usepackage{ragged2e}
\usepackage{setspace}
\usepackage{verbatim}
\usepackage{comment}
\usepackage{microtype}
\usepackage{afterpage}
\usepackage{soul}
\usepackage{ulem}
\usepackage{enumitem}
\usepackage{needspace}
\usepackage{adjustbox}
\usepackage{float}
\usepackage{multirow}
\usepackage{booktabs}
\usepackage{tabularx}
\usepackage{hhline}
\usepackage{simpler-wick}
\usepackage{braket}
\newcommand{\hv}{\langle h\rangle}

\usepackage{hyperref}
\usepackage[all]{hypcap}
\hypersetup{
	colorlinks=true,
	linkcolor={linkcolor},
	citecolor={linkcolor},
	urlcolor={linkcolor},
	linktocpage=true
}
\definecolor{linkcolor}{rgb}{0.0, 0.28, 0.67}

\usepackage{caption}
\DeclareCaptionJustification{justified}{\justifying}
\usepackage{tcolorbox}
\usepackage{pifont}
\usepackage{fontawesome}
\usepackage{orcidlink}
\usepackage{framed}

\definecolor{oucrimsonred}{rgb}{0.6, 0.0, 0.0}
\definecolor{persianblue}{rgb}{0.11, 0.22, 0.73}
\definecolor{forestgreen}{rgb}{0.13,0.35,0.13}
\definecolor{harvardcrimson}{rgb}{0.79, 0.0, 0.09}
\definecolor{oceanboatblue}{rgb}{0.0, 0.47, 0.75}
\definecolor{egyptianblue}{rgb}{0.06, 0.2, 0.65}
\definecolor{navyblue}{rgb}{0.0, 0.0, 0.5}
\definecolor{verdechiaro}{rgb}{0.6,1,0.6}
\definecolor{giallochiaro}{rgb}{1,1,0.6}
\definecolor{bluscuro}{rgb}{0.15, 0.2, 0.9}
\definecolor{verdes}{rgb}{0.1, 0.5, 0.1}
\definecolor{tangerineyellow}{rgb}{1.0, 0.8, 0.0}
\definecolor{azure}{rgb}{0.0, 0.5, 1.0}
\definecolor{deepfuchsia}{rgb}{0.76, 0.33, 0.76}
\definecolor{VioletRed4}{rgb}{0.55, 0.13, .32}
\definecolor{colorRTD}{rgb}{.2,.2,.7}
\definecolor{colorRTD2}{rgb}{.2,.3,.7}
\definecolor{bluchiaro}{cmyk}{1,.3,0.,0.1}
\definecolor{ForestGreen}{rgb}{0.13, 0.55, 0.13}
\definecolor{rossos}{cmyk}{0,1,1,0.55}

\newcommand{\Loop}{(16\pi^2)}
\newcommand{\mO}{\mathcal{O}}

\newcommand{\Lh}{{\Lambda}_{H}}

\newcommand{\nn}{\nonumber}

\def\nn{\nonumber}

\def\lsim{\mathrel{\rlap{\lower4pt\hbox{\hskip0.5pt$\sim$}} \raise1pt\hbox{$<$}}}
\def\gsim{\mathrel{\rlap{\lower4pt\hbox{\hskip0.5pt$\sim$}} \raise1pt\hbox{$>$}}}

\newcommand{\be}{\begin{eqnarray}}
	\newcommand{\ee}{\end{eqnarray}}
\newcommand{\bs}{\begin{subequations}}
	\newcommand{\es}{\end{subequations}}

\newtcolorbox{mybox}{colback=mycolor!5!white,colframe=azure!75!black}

\begin{document}

\title{Weak Scale Triggers in the SMEFT}

\author[a,b]{Pier~Giuseppe~Catinari\orcidlink{0009-0008-5416-1770}}
\affiliation[a]{Dipartimento di Fisica, La Sapienza Università di Roma, Piazzale Aldo Moro 2, I-00185 Rome, Italy}
\affiliation[b]{INFN Sezione di Roma, Piazzale Aldo Moro 2, I-00185 Rome, Italy}

\author[c,d]{Raffaele~Tito~D'Agnolo\orcidlink{https://orcid.org/0000-0002-8271-6840}}
\affiliation[c]{Institut de Physique Th\'eorique, Universit\'e Paris-Saclay, CEA, F-91191 Gif-sur-Yvette, France}
\affiliation[d]{Laboratoire de Physique de l’École Normale Supérieure, ENS, Université PSL, CNRS, Sorbonne
Université, Université Paris Cité, F-75005 Paris, France}

\author[e]{Pablo~Sesma\orcidlink{https://orcid.org/0000-0002-9035-3322}}
\affiliation[e]{Institut de Física d’Altes Energies (IFAE) and The Barcelona Institute of Science
and Technology (BIST), Campus UAB, 08193 Bellaterra, Barcelona, Spain}

\emailAdd{piergiuseppe.catinari@uniroma1.it}
\emailAdd{raffaele-tito.dagnolo@ipht.fr}
\emailAdd{psesma@ifae.es}

\abstract{
There are no weak scale triggers in the SMEFT up to dimension six that can solve the hierarchy problem far above the weak scale. Our arguments can be used to show that the same is true  at dimension eight. Weak scale triggers are local operators sensitive to the Higgs mass squared and they are needed in a large number of qualitatively different cosmological solutions to the hierarchy problem. These solutions have little in common besides the use of a trigger operator. We argue that focusing on the signatures of the three already-known trigger operators can lead to discover or exclude this class of solutions to the hierarchy problem.
}
\maketitle

\section{Introduction}
Weak scale triggers are local operators $\mathcal{O}_T$ whose vacuum expectation value (VEV) depends on the Higgs boson mass squared. More precisely, we require the VEV of $\mathcal{O}_T$ to change at $O(1)$ for a variation of $m_h^2$ of the same magnitude,
\be\label{eq:Trigger definition}
\frac{d \log \langle \mathcal{O}_T \rangle}{d\log m_h^2} = O(1)\, . \label{eq:TDef}
\ee
These objects answer a basic quantum field theory question: is there anything that changes (locally) in Nature when we change the Higgs mass squared? Within the Standard Model (SM) the answer is just one operator\footnote{However there are plenty of non-local objects with the same property. For instance, all two-point functions of massive fermions.}, and we discuss it in Section~\ref{sec:existing}. 

Triggers are relevant to the Higgs hierarchy problem and to electroweak symmetry breaking in general.
They were first explicitly discussed in~\cite{Arkani-Hamed:2020yna}, but they had already been used in multiple ways to solve the Higgs hierarchy problem~\cite{Graham:2015cka, Espinosa:2015eda, Geller:2018xvz,  Cheung:2018xnu,  Arkani-Hamed:2020yna, Strumia:2020bdy, Csaki:2020zqz, TitoDAgnolo:2021nhd, TitoDAgnolo:2021pjo,  Chatrchyan:2022pcb, Trifinopoulos:2022tfx, Csaki:2022zbc, Matsedonskyi:2023tca, Hook:2023yzd, Chattopadhyay:2024rha, Csaki:2024ywk, Chatrchyan:2022dpy}. 
Identifying triggers and studying their phenomenology allows to probe model-independently a large class of solutions to the problem~\cite{TitoDAgnolo:2021pjo}. Discovering that the observed Higgs mass is selected by a trigger operator would cause an epistemological shift in our understanding of the electroweak hierarchy problem.
Symmetries that solve the hierarchy problem make $m_h^2 \simeq 0$ special in the fundamental theory of Nature. Trigger operators make $m_h^2 \simeq 0$ special for the evolution of our universe. In theories with triggers, any value of $m_h^2$ is permitted and $m_h^2 \simeq M_{\rm Pl}^2$ remains the most natural possibility\footnote{Here and in the following we assume $O(1)$ couplings at the Planck scale, so that $M_{\rm Pl}\simeq 10^{19}$~GeV.}, but our particular universe sees a value $m_h^2 \ll M_{\rm Pl}^2$ due to its cosmological evolution. Conceptually, explaining the value of $m_h^2$ is demoted from understanding a fundamental parameter of Nature to justifying an accident occurring in our universe. Any experimental evidence for this new paradigm would fundamentally alter the way we understand the SM and its other fine-tuning problems.

In this work we try to identify new triggers by considering operators of dimension 5 and 6 in the Standard Model Effective Field Theory (SMEFT) and find no new viable candidate. Our arguments lead to the same conclusion for operators of dimension 8~\cite{Li:2020gnx}, but here we show a single explicit example.   

Studying the SMEFT allows to make model-independent statements on triggers beyond the SM. Even if we do not prove any theorem, {\it de facto} our study implies that the only viable triggers are the three examples that are already well-known (one in the SM plus two BSM triggers) and we discuss in Section~\ref{sec:existing}. This result is important because it allows to identify, {\it ex negativo}, the experimental signatures needed to prove or falsify the trigger paradigm and with it a vast class of solutions to the hierarchy problem~\cite{Graham:2015cka, Espinosa:2015eda, Geller:2018xvz,  Cheung:2018xnu, Arkani-Hamed:2020yna, Strumia:2020bdy, Csaki:2020zqz, TitoDAgnolo:2021nhd, TitoDAgnolo:2021pjo,  Chatrchyan:2022pcb, Trifinopoulos:2022tfx, Csaki:2022zbc, Matsedonskyi:2023tca, Hook:2023yzd, Chattopadhyay:2024rha, Csaki:2024ywk, Chatrchyan:2022dpy}.
Interestingly, most of these signatures are well within reach of the next two decades of axion and high-energy experimental programs~\cite{TitoDAgnolo:2021pjo}.

To be more precise on the nature of our results, we recall that the VEV of most operators in the SM and the SMEFT is UV-sensitive and receives contributions proportional to some powers of the cutoff of the theory\footnote{When we make statements about the cutoff we always make an implicit assumption on the UV. We assume that above $\Lambda$ a UV-theory where the operator VEV is calculable exists. So our cutoff is a physical Wilsonian cutoff, as for example the mass of MSSM superpartners.}, $\langle \mathcal{O} \rangle \simeq c_1\Lambda^n/16\pi^2$. Some of these operators also receive tree-level contributions proportional to the Higgs VEV $\langle \mathcal{O} \rangle \simeq c_2 v^n$. In practice Eq.~\eqref{eq:TDef} can only be satisfied up to some value of the cutoff $\Lambda \lesssim (16 \pi^2 c_2/c_1)^{1/n}$. Therefore an operator can be a trigger in a given EFT with a given cutoff $\Lambda_1$, but not be a trigger in the same EFT with a larger cutoff $\Lambda_2 > \Lambda_1$\footnote{Or in a different EFT with the same cutoff. We  discuss this possibility in the following.}.

Our study of the SMEFT estimates upper bounds on $\Lambda$ for all possible operators of a given dimension. We find no viable operator with $\Lambda > 4\pi v$ up to dimension 8 (although here we show explicit results only up to dimension 6 and a single example at dimension 8). We do not find particularly compelling to solve the hierarchy problem only up to $4\pi v$ by invoking the existence of triggers and the related cosmological dynamics, since this is the scale where we would already expect new states from symmetry-based solutions to the problem. This means that new trigger operators can only exist in BSM theories with new physics below $4\pi v$ or at higher operator dimension\footnote{A bound on $\Lambda$ requires computing the VEV of the corresponding operator. In general this is a multi-loop calculation already at dimension 6. In this work we give all the parametric scalings of the result, but we do not compute all $O(1)$ coefficients. Therefore we cannot exclude that for some operators the cutoff is somewhat larger than $4\pi v$. As we will see in the following this is unlikely to affect our conclusion, since going a factor of 10 above $4\pi v$ requires coefficients that should naively be $O(1)$ to be around $10^4$. We give more details on this point in Section~\ref{sec:SMEFT}.}. Going to high operator dimensions presents severe model-building challenges when trying to justify the existence of the trigger and its role in solving the hierarchy problem, as we discuss in the following. Introducing new physics below $4\pi v$, with the additional requirement of generating a trigger, is quite challenging after more than 10 years of running of the LHC, and only two viable examples of BSM triggers exist~\cite{Arkani-Hamed:2020yna, TitoDAgnolo:2021pjo}. Therefore our results imply that the three known trigger operators are by far the most interesting to focus on experimentally and that it is highly unlikely that other viable operators exist. 

The rest of the paper is organized as follows. In Section~\ref{sec:existing} we discuss in more detail what is a trigger, which triggers exist in the SM and how they can be used to solve the hierarchy problem. In Section~\ref{sec:general} we show how to compute the VEV of a trigger and how to divide SMEFT operators based on a symmetry principle that allows to see immediately if they can be triggers or not. In Section~\ref{sec:SMEFT} we give our results for SMEFT operators up to dimension 6. In Section~\ref{sec:noT} we review two operators that fail to be triggers for non-trivial reasons (one of them is actually a trigger in an appropriately chosen BSM theory, but we do not know how to use it to solve the hierarchy problem). We conclude in Section~\ref{sec:conclusion} summarizing the implications of our results.

\section{Existing Triggers and Their Applications}\label{sec:existing}

At present only three trigger operators are known. One exists in the SM and two more require new states charged under $SU(2)_L\times U(1)_Y$ to lie in the few hundreds of GeV range. 

The only SM trigger is the antisymmetric contraction of two gluon field strengths
\be
G\widetilde G \equiv \frac{1}{2} \sum_{a=1}^8 \epsilon^{\mu\nu\rho\sigma} G_{\mu\nu}^a G_{\rho \sigma}^a\, , \label{eq:GGdual}
\ee
which is the operator associated to the strong CP problem. $G\widetilde G$ can be written as the total derivative of a current $\partial_\mu K^\mu$. The symmetry associated to $K^\mu$ is broken by non-perturbative effects of order $e^{-8\pi^2/g_s^2}$, where $g_s$ is the coupling constant of strong interactions. At high energies $g_s$ runs to zero in the SM. Therefore the VEV of $G\widetilde G$ is dominated by energy scales where $g_s$ is large and is shielded from high energy contributions at the scale\footnote{From now on we call the cutoff of the theory $\Lambda_H$ because we assume that a symmetry that solves the hierarchy problem and makes all VEVs calculable exists at this scale.} $\Lambda_H$ by the exponential factor $e^{-8\pi^2/g_s^2(\Lambda_H)}$. Below the QCD scale one can compute $\langle G\widetilde G \rangle$ from the chiral Lagrangian and obtain $\langle G\widetilde G \rangle \sim \bar{\theta} m_\pi^2 f_\pi^2 \zeta$, where $\zeta=m_u m_d/(m_u+m_d)^2$. One can obtain the same result from current algebra~\cite{Shifman:1979if}. Above the QCD scale, where $g_s$ becomes perturbative, we can perform an instanton calculation and get a much smaller result  $\langle G\widetilde G \rangle \simeq \bar{\theta} m_\pi^2 f_\pi^2 \zeta +O(e^{-8\pi^2/g_s^2(\Lambda_H)}\Lambda_H^4)$. Overall the VEV is dominated by contributions at the QCD scale and is protected from high energy contributions,
\be
\langle G\widetilde G \rangle \sim \bar{\theta} m_\pi^2 f_\pi^2 \frac{m_u m_d}{(m_u+m_d)^2} \gg e^{-8\pi^2/g_s^2(\Lambda_H)}\Lambda_H^4\, ,
\ee
thanks to the smallness of the strong coupling at high energies.
As a consequence, the VEV of $G\widetilde G$ is proportional to the Higgs VEV squared
\be
\langle G\widetilde G \rangle \sim \bar{\theta} m_\pi^2 f_\pi^2 \frac{m_u m_d}{(m_u+m_d)^2} \sim v^2\, ,
\ee
even when $\Lambda_H \simeq M_{\rm Pl}$. One power of $v$ comes from quark masses in $m_\pi$, the second power from the running of $g_s$. This fact has already been exploited to propose many qualitatively different solutions to the hierarchy problem, starting with~\cite{Graham:2015cka}, a joint solution to the hierarchy problem and the strong CP problem~\cite{TitoDAgnolo:2021nhd} and a solution to the doublet-triplet splitting problem in GUTs~\cite{Csaki:2024ywk}. All other operators in the SM have VEVs sensitive to the cutoff of the theory and are not strongly affected by changes in $m_h^2$. For example $\langle |H|^2 \rangle \simeq v^2 + \Lambda_H^2/16\pi^2$ and $d\log \langle |H|^2 \rangle/d \log m_h^2 \sim 16 \pi^2 m_h^2 /\Lambda_H^2 \ll 1$ if $\Lambda_H \gg m_h$. We discuss other SM operators in the next Section.

Naively one might think that infinitely many triggers exist beyond the SM and on paper this is true. To write down a trigger operator it is sufficient to find a symmetry that protects $\langle \mathcal{O}_T \rangle$ which is only broken by the Higgs VEV. Overall it is a rather simple model-building exercise. However we cannot introduce in our BSM theory any mass scale larger than $m_h$. Any scale $M$ that contributes to $\langle \mathcal{O}_T \rangle$ enters our Eq.~\eqref{eq:TDef} as
\be
\frac{d \log \langle \mathcal{O}_T \rangle}{d\log m_h^2}\propto \left(\frac{m_h^2}{M^2}\right)^n\, , \quad n >0\, , \label{eq:cutoff}
\ee
potentially making the derivative much smaller than 1. If $m_h^2/M^2 \ll 1$ changing $m_h^2$ barely affects $\langle \mathcal{O}_T \rangle$ and the evolution of the Universe becomes effectively insensitive to $m_h^2$. The requirement $d \log \langle \mathcal{O}_T \rangle/d\log m_h^2 \sim 1$ is thus more stringent than the traditional naturalness requirement of having new states one loop above $m_h^2$ and makes it almost impossible to find any trigger beyond the SM consistent with experiment. 

To better understand why we need new physics lighter than $m_h$, it is useful to briefly describe how a trigger can solve the hierarchy problem. 
We can use $\mathcal{O}_T$ to solve the hierarchy problem by coupling it to one (or many) new singlet scalars $\phi$ that play a role in the evolution of the Universe. If Eq.~\eqref{eq:TDef} is satisfied, an $O(1)$ variation of $m_h^2$ can affect the $\phi$ potential at $O(1)$ and couple the evolution of the Universe to the value of $m_h^2$. If paired with a mechanism to populate multiple values of $m_h^2$, this can explain the unnaturally small observed value of the Higgs mass. Examples include crunching all patches of the Multiverse with the ``wrong'' Higgs mass~\cite{TitoDAgnolo:2021nhd, TitoDAgnolo:2021pjo, Csaki:2020zqz, Csaki:2024ywk}, generating a small cosmological constant only in patches where $m_h^2$ is also small~\cite{Arkani-Hamed:2020yna}, stopping dynamically the rolling of $\phi$ where $m_h^2$ is small~\cite{Graham:2015cka}, inflating for longer times in patches with small values of $m_h^2$~\cite{Geller:2018xvz, Giudice:2021viw}, and many others. 

One might argue that nothing stops a creative model-builder from engineering a $\phi$ potential that is strongly affected by a small relative variation of $\langle \mathcal{O}_T \rangle$. However this is precisely the definition of fine tuning and would just move the hierarchy problem from $m_h^2$ to the $\phi$ potential. 

Technically, also the cosmological solutions to the hierarchy problem listed above move the problem from the Higgs potential to the $\phi$ potential $V_\phi$, as we still need a small number in $V_\phi$ that explains the hierarchy $m_h^2/\Lambda_H^2 \ll 1$. However they do it in a technically natural way. Those ideas have a $\phi$ sector where an approximate symmetry (usually supersymmetry or scale invariance) protects a large hierarchy of scales, and allows $V_\phi$ to be naturally much smaller than $\Lambda_H^4$. A small Higgs mass is then selected in our sector via a weak coupling between $\phi$ and $\mathcal{O}_T$ that affects at $O(1)$ the small $V_\phi$ when $-m_h^2 \simeq (125\,\text{GeV})^2 \ll \Lambda_H^2$. The extremely weak coupling of $\phi$ to the SM avoids that the breaking of supersymmetry (or scale invariance) in our sector affects $V_\phi$. To make this discussion clearer, we can consider the following toy example
\be
V = V_\phi + \phi \mathcal{O}_T\, .
\ee
If in a typical point in field space $V_\phi \simeq \langle \phi \rangle \langle \mathcal{O}_T \rangle_{m_h^2=-(125\,{\rm GeV})^2}$ and $\langle \mathcal{O}_T \rangle$ is a monotonic function of $m_h^2$ that satisfies Eq.~\eqref{eq:TDef} we can have a technically natural solution of the hierarchy problem, provided that we can write down a symmetry that makes $V_\phi \ll \Lambda_H^4$. We see immediately that $m_h^2=-(125\,{\rm GeV})^2$ is a special point in this theory. When we move away from it, $\langle \phi \rangle \langle \mathcal{O}_T \rangle$ becomes larger or smaller than $V_\phi$. Now imagine instead that $\langle \mathcal{O}_T \rangle \simeq M^3 + m_h^3$ with $M^3 \gg (125\,{\rm GeV})^3$. If $V_\phi \simeq \langle \phi \rangle \langle \mathcal{O}_T \rangle_{m_h^2=-(125\,{\rm GeV})^2}$ then the $\phi$ dynamics is dominated by $M$ when $m_h$ is close to the weak scale and knows nothing about $m_h$. If instead $V_\phi \simeq \langle \phi \rangle (125\,{\rm GeV})^3$ we are still not selecting the observed Higgs mass because $V_\phi \ll \langle \phi \rangle \langle \mathcal{O}_T \rangle$ for a large range of values of $m_h^2$, including the measured one. The only option is to sit on a point in field space where a small relative variation of the potential (of order $(m_h/M)^3$) changes significantly the dynamics of $\phi$, but this is precisely a tuning of initial conditions.

The above discussion highlights the difficulties of finding triggers beyond the SM. However, two examples exist and they are 1) $H_1 H_2$, where $H_1$ is a new Higgs doublet and the VEV of $H_1 H_2$ is protected by a discrete symmetry~\cite{Arkani-Hamed:2020yna} and 2) $F\widetilde F$ where $F$ is the field strength of a new confining gauge group. In this second case one needs to introduce new vector-like leptons charged under the group (minimally an $SU(2)_L$ doublet $L$ and a singlet $N^c$) which get an $O(1)$ fraction of their mass from the Higgs VEV~\cite{Graham:2015cka}. Both of these triggers offer exciting detection opportunities at HL-LHC and future lepton colliders~\cite{Arkani-Hamed:2020yna, Beauchesne:2017ukw, TitoDAgnolo:2021pjo}.

It is hard to prove in full generality that no other BSM triggers can exist other than the two examples above, but writing a new trigger requires new states charged under $SU(2)_L \times U(1)_Y$ at or below $m_h$. The two existing examples already contain the smallest possible $SU(2)_L$ representations and are close to being excluded by the LHC~\cite{Arkani-Hamed:2020yna, Beauchesne:2017ukw}. Anything more complicated, i.e. containing larger SM representations or more new states, is bound to be already experimentally excluded. Therefore we find more fruitful to ask if one can find new trigger operators using only the fields that we have already discovered. The rich symmetry structure of the SM, which is full of approximate global symmetries only broken by small parameters, is a good place to look for new triggers. We can see this as follows.

Let $d$ be the operator dimension. In an EFT completed into a UV theory where the VEV is calculable at a scale $\Lambda_H$, all contributions to $\langle\mathcal{O}\rangle$ are of the form
\begin{equation}
    \label{eq:protoT}\langle\mathcal{O}\rangle =\sum_i \frac{\varepsilon_i}{(16\pi^2)^{\ell_i}}\Lambda_H^{d-n_i}\hv^{n_i} \, ,
\end{equation}
where $\ell_i$ counts loops, $n_i\geq 0$ is the power of $\hv$ enforced by symmetries, and $\varepsilon_i$ are hard symmetry-breaking spurions (Yukawas, CKM matrix elements, phases, gauge couplings, etc.). Note that we call $v$ the Higgs VEV in our universe, $v\simeq 174$~GeV, while $\hv$ is a generic Higgs VEV that can be very different from $v$, either because cosmological dynamics has not yet relaxed it to its SM value or because we are talking about a different universe in the Multiverse. From Eq.~\eqref{eq:protoT} we conclude that
\begin{equation}
    \frac{d\log\langle\mathcal{O}\rangle}{d\log m_h^2}=\frac{\sum_i w_i \frac{n_i}{2}}{\sum_i w_i}\, , \quad w_i\equiv \frac{\varepsilon_i}{(16\pi^2)^{\ell_i}}\Lambda_H^{d-n_i}\hv^{n_i}\, .
\end{equation}
As $\Lambda_H\rightarrow \infty$, terms with the smallest $n_i$ (fewest powers of $\hv$) dominate. Therefore, a necessary and sufficient condition for a trigger that solves more than the little hierarchy problem ($\Lambda_H \lesssim 4\pi v$) is that all $\hv$-independent terms are suppressed by (approximate) symmetries, i.e. $\epsilon_i \lesssim (v/\Lambda_H)^{d-n_i}$ for all $d-n_i>0$, because the relevant value of $\hv$ for an upper bound on $\Lambda_H$ is $\hv=v$. Since we have to make our universe special, we need $\langle \mathcal{O}\rangle$ to be dominated by contributions proportional to $\hv$ when the Higgs VEV is at least the size in our Universe, so we rarely refer to $\hv$ again in what follows.

In this work we compute Eq.~\eqref{eq:protoT} for all $d=5,6$ operators in the SMEFT and set an upper bound on $\Lambda_H$. In doing so we always set $\hv=v$, as we did for $\epsilon_i$ above.

\section{General Strategy and Selected Examples}\label{sec:general}
To find trigger operators in the SMEFT we systematically compute their VEV, using the following simple technique, already described in~\cite{Arkani-Hamed:2020yna}. We introduce in the theory a probe scalar $\phi$ and imagine that $\phi$ is the lightest particle in the spectrum. We give $\phi$ a parametrically weak coupling $\xi$ to the operator of interest $\mathcal{O}_{\rm SM}$, and no other interaction,
\be
\mathcal{L}(\Phi,\phi) = \frac{(\partial \phi)^2}{2} - \xi \phi \mathcal{O}_{\rm SM}(\Phi)+\mathcal{L}_{\rm SM}(\Phi)\, ,
\ee
where $\Phi$ collects all SM fields. We integrate out all the fields in $\mathcal{L}_{\rm SM}$, then expand the resulting low-energy $\phi$ potential for small $\xi$ and obtain
\be
\mathcal{L}_{\rm EFT} &=& \frac{(\partial \phi)^2}{2} -V_{\rm EFT}(\xi \phi) \nn \\
V_{\rm EFT}(\xi \phi) &=& \xi \phi \langle \mathcal{O}_{\rm SM} \rangle + (\xi \phi)^2 + \ldots\, . 
\ee
The first term in the expansion is proportional to the VEV of $\mathcal{O}_{\rm SM}$, so we can simply extract $\langle \mathcal{O}_{\rm SM} \rangle$ from
\be
\langle \mathcal{O}_{\rm SM} \rangle = \left.\frac{d V_{\rm EFT}(x)}{d x}\right|_{x=0}\, . \label{eq:Tvev}
\ee
Note that this is no different from writing the generating functional of the theory,
\begin{equation}
    Z_{\rm SM}[\lambda]=\int\mathcal{D}\Phi~\exp\left\{i\int d^4x~\left(\mathcal{L}_{\rm SM}+\lambda\, \mathcal{O}_{\rm SM} \right) \right\}\,, \quad W_{\rm SM}[\lambda]=-i\log Z_{\rm SM}[\lambda]\, ,
\end{equation}
promoting $\lambda$ to a background field, and computing the VEV in the theory deformed by $\lambda$ as
\begin{equation}
    \langle \mathcal{O}_{\rm SM}(x)\rangle_{\lambda}=\frac{1}{Z_{\rm SM}[\lambda]}\int\mathcal{D}\Phi~\mathcal{O}_{\rm SM}(x) e^{iS[\Phi;\,\lambda]}=\frac{\delta W_{\rm SM}}{\delta\lambda(x)}\, .
\end{equation}
Setting $\lambda=0$ then gives the vacuum expectation value in the undeformed SM,
\begin{equation}
    \langle \mathcal{O}_{\rm SM}(x)\rangle=\left.\frac{\delta W_{\rm SM}}{\delta\lambda(x)}\right|_{\lambda=0}\,.
\end{equation}
In our theory, $\lambda$ is the background determined by the light field, i.e. $\lambda(x)=-\xi\, \phi(x)$. Integrating out $\Phi$ first gives
\begin{align}
    Z&=\int \mathcal{D}\phi~\exp\left\{i\int d^4x~\frac{1}{2}(\partial\phi)^2 \right\}Z_{\rm SM}[\lambda=-\xi \phi]\nn\\&=\int \mathcal{D}\phi~\exp\Big\{i\,  S_0[\phi]+i\,W_{\rm SM}[\lambda=-\xi\phi] \Big\}\, .
\end{align}
Therefore the (quantum) effective action for $\phi$ after integrating out the SM is $\Gamma[\phi]=S_0[\phi]+W_{\rm SM}[\lambda=-\xi\phi]$ up to corrections from $\phi$-loops. If $\phi$ is weakly coupled and we are only interested in triggers, we can work at leading order in $\phi$-loops, i.e. take
\begin{equation}
    \Gamma[\phi]\simeq \int d^4x~\frac{1}{2}(\partial\phi)^2+W_{\rm SM}[\lambda=-\xi\phi]\, .
\end{equation}
Treating $\phi$ as a constant background \textit{à la} Coleman-Weinberg~\cite{Coleman:1973jx} we have
\begin{equation}
    \Gamma[\phi(x)=\phi]=-\int d^4x~V_{\rm eff}(\phi)\, ,
\end{equation}
from which we immediately obtain 
\begin{equation}
    V_{\rm eff}(\phi)=-\frac{1}{\mathcal{V}_4}W_{\rm SM}[\lambda=-\xi\phi]\, ,
\end{equation}
where $\mathcal{V}_4$ is the volume of spacetime.
Differentiating with respect to $\phi$, we get
\begin{equation}
    \frac{dV_{\rm eff}}{d\phi}=-\frac{1}{\mathcal{V}_4}\int d^4x~\frac{\delta W_{\rm SM}}{\delta\lambda(x)}\frac{\delta \lambda(x)}{\delta\phi}=-\frac{1}{\mathcal{V}_4}\int d^4x~\langle\mathcal{O}_{\rm SM}(x)\rangle_{\lambda}(-\xi)=\xi\langle\mathcal{O}_{\rm SM}\rangle_{\lambda}\, .
\end{equation}
At $\phi=0$, i.e. $\lambda=0$, this becomes the same as Eq.~\eqref{eq:Tvev},
\begin{equation}
    \left.\frac{d V_{\rm eff}}{d\phi}\right|_{\phi=0}=\xi \langle \mathcal{O}_{\rm SM}\rangle\, .
\end{equation}
We now systematically analyze all the independent dimension 6 operators $\mathcal{O}^{(6)}_k$ in~\cite{Grzadkowski:2010es} and the dimension-five operator $\mathcal{O}^{(5)}=(HL)^2$~\cite{Weinberg:1979sa}, where
\begin{align}   
\mathcal{L}_{\rm{SMEFT}}&= \mathcal{L}_{\rm{SM}}^{(d\leq 4)}+\frac{1}{M_{\rm UV}} C^{(5)}\mathcal{O}^{(5)}+\frac{1}{M_{\rm UV}^2}\sum_k C_k^{(6)}\mathcal{O}^{(6)}_k+\sum_{d\geq 7}\frac{1}{M_{\rm UV}^{d-4}}\sum_{i}C_i^{(d)}\mathcal{O}_i^{(d)}\,.
\end{align}
We take $M_{\rm UV}$ slightly above $\Lambda_H$, as discussed in the next Section.

\subsection{Selected Examples}\label{sec:examples}

In this Section we start by giving a few examples that cover most qualitatively distinct cases. We are going to use two-component spinor notation where the SM fermion fields and their quantum numbers are 
\be
\Psi =\{Q(3,2)_{1/6}, L(1,2)_{-1/2}, u^c(3,1)_{-2/3}, d^c(3,1)_{1/3}, e^c(1,1)_1, N(1,1)_0\}\, .
\ee
We introduce the following shorthand notation for a generic $SU(2)_L$ doublet or singlet
\be
\psi=\{Q, L\}\, , \quad \psi^c=\{u^c, d^c, e^c, N\}\, , \quad q^c=\{u^c, d^c\}\, ,
\ee
and we use the following convention for the Higgs field
\be
H=(1,2)_{1/2}\, , \quad \tilde H = i \sigma^2 H^*\, .
\ee
Latin indices are for flavor and, when useful, we write the field explicitly (for instance $u^{c\dagger} \bar\sigma^\mu c$ is a contraction of up and charm quarks).

\begin{figure}
    \centering

\tikzset{every picture/.style={line width=1pt}} 

\begin{tikzpicture}[x=1pt,y=1pt,yscale=-1,xscale=1]

\draw  [fill={rgb, 255:red, 0; green, 0; blue, 0 }  ,fill opacity=1 ] (404.13,91.63) -- (409.03,91.63) -- (409.03,96.53) -- (404.13,96.53) -- cycle ;
\draw  [dash pattern={on 0.84pt off 2.51pt}]  (406.58,94.08) -- (429.94,117.08) ;
\draw  [dash pattern={on 0.84pt off 2.51pt}]  (383.23,117.08) -- (406.57,94.1) ;
\draw  [dash pattern={on 0.84pt off 2.51pt}] (428.25,69.49) .. controls (430.12,71.14) and (426.78,77.98) .. (420.79,84.77) .. controls (414.8,91.57) and (408.43,95.74) .. (406.57,94.1) .. controls (404.7,92.46) and (408.05,85.62) .. (414.03,78.82) .. controls (420.02,72.03) and (426.39,67.85) .. (428.25,69.49) -- cycle ;
\draw  [dash pattern={on 0.84pt off 2.51pt}] (382.3,70.34) .. controls (384.04,68.57) and (390.7,72.26) .. (397.17,78.6) .. controls (403.64,84.94) and (407.48,91.51) .. (405.74,93.29) .. controls (404,95.06) and (397.35,91.37) .. (390.87,85.03) .. controls (384.4,78.7) and (380.56,72.12) .. (382.3,70.34) -- cycle ;
\draw  [fill={rgb, 255:red, 0; green, 0; blue, 0 }  ,fill opacity=1 ] (317.7,91.32) -- (322.6,91.32) -- (322.6,96.22) -- (317.7,96.22) -- cycle ;
\draw  [dash pattern={on 0.84pt off 2.51pt}]  (296.61,70.46) -- (343.32,116.46) ;
\draw  [dash pattern={on 0.84pt off 2.51pt}] (341.64,68.87) .. controls (343.5,70.52) and (340.16,77.36) .. (334.17,84.15) .. controls (328.18,90.95) and (321.81,95.13) .. (319.95,93.48) .. controls (318.08,91.84) and (321.43,85) .. (327.42,78.2) .. controls (333.4,71.41) and (339.77,67.23) .. (341.64,68.87) -- cycle ;

\draw  [dash pattern={on 0.84pt off 2.51pt}]  (296.61,116.46) -- (319.95,93.48) ;

\draw  [fill={rgb, 255:red, 0; green, 0; blue, 0 }  ,fill opacity=1 ] (233.8,91.97) -- (238.7,91.97) -- (238.7,96.87) -- (233.8,96.87) -- cycle ;
\draw  [dash pattern={on 0.84pt off 2.51pt}]  (212.9,71.42) -- (259.6,117.42) ;
\draw  [dash pattern={on 0.84pt off 2.51pt}]  (213.11,117.29) -- (259.39,71.54) ;
\draw  [dash pattern={on 0.84pt off 2.51pt}]  (236.29,126.13) -- (236.25,62.12) ;
\draw  [fill={rgb, 255:red, 0; green, 0; blue, 0 }  ,fill opacity=1 ] (491.47,92.3) -- (496.37,92.3) -- (496.37,97.2) -- (491.47,97.2) -- cycle ;
\draw  [dash pattern={on 0.84pt off 2.51pt}] (515.6,70.14) .. controls (517.47,71.79) and (514.12,78.63) .. (508.14,85.42) .. controls (502.15,92.22) and (495.78,96.39) .. (493.92,94.75) .. controls (492.05,93.11) and (495.4,86.27) .. (501.38,79.47) .. controls (507.37,72.68) and (513.74,68.5) .. (515.6,70.14) -- cycle ;
\draw  [dash pattern={on 0.84pt off 2.51pt}] (469.26,70.93) .. controls (471,69.15) and (477.65,72.85) .. (484.13,79.19) .. controls (490.6,85.52) and (494.44,92.1) .. (492.7,93.87) .. controls (490.96,95.65) and (484.3,91.95) .. (477.83,85.62) .. controls (471.36,79.28) and (467.52,72.7) .. (469.26,70.93) -- cycle ;
\draw  [dash pattern={on 0.84pt off 2.51pt}] (493.7,94.87) .. controls (496.18,94.84) and (498.29,102.16) .. (498.41,111.21) .. controls (498.53,120.27) and (496.61,127.64) .. (494.12,127.67) .. controls (491.64,127.7) and (489.53,120.39) .. (489.41,111.33) .. controls (489.29,102.27) and (491.21,94.91) .. (493.7,94.87) -- cycle ;
\draw  [dash pattern={on 0.84pt off 2.51pt}]  (259.49,37.49) -- (213.01,37.37) ;
\draw  [fill={rgb, 255:red, 0; green, 0; blue, 0 }  ,fill opacity=1 ] (317.74,35.4) -- (322.64,35.4) -- (322.64,40.3) -- (317.74,40.3) -- cycle ;
\draw  [dash pattern={on 0.84pt off 2.51pt}] (305.29,22.95) .. controls (305.29,14.72) and (311.96,8.05) .. (320.19,8.05) .. controls (328.41,8.05) and (335.09,14.72) .. (335.09,22.95) .. controls (335.09,31.18) and (328.41,37.85) .. (320.19,37.85) .. controls (311.96,37.85) and (305.29,31.18) .. (305.29,22.95) -- cycle ;
\draw  [fill={rgb, 255:red, 0; green, 0; blue, 0 }  ,fill opacity=1 ] (233.8,34.98) -- (238.7,34.98) -- (238.7,39.88) -- (233.8,39.88) -- cycle ;
\draw  [dash pattern={on 0.84pt off 2.51pt}]  (319.77,126.1) -- (320.15,93.77) ;

\draw (150.83,88.5) node [anchor=north west][inner sep=0.75pt]    {$\langle |H|^{6} \rangle \simeq $};
\draw (274.83,88.5) node [anchor=north west][inner sep=0.75pt]    {$+$};
\draw (359.33,88.5) node [anchor=north west][inner sep=0.75pt]    {$+$};
\draw (450.33,88.5) node [anchor=north west][inner sep=0.75pt]    {$+$};
\draw (520.33,88.5) node [anchor=north west][inner sep=0.75pt]    {$+\dots $};
\draw (150.83,30.95) node [anchor=north west][inner sep=0.75pt]    {$\langle |H|^{2} \rangle \simeq $};
\draw (275.17,32.95) node [anchor=north west][inner sep=0.75pt]    {$+$};
\draw (345.17,34.95) node [anchor=north west][inner sep=0.75pt]    {$,$};

\draw (429.94-3pt,117.08-3pt) -- (429.94+3pt,117.08+3pt);
\draw (429.94-3pt,117.08+3pt) -- (429.94+3pt,117.08-3pt);

\draw (383.23-3pt,117.08-3pt) -- (383.23+3pt,117.08+3pt);
\draw (383.23-3pt,117.08+3pt) -- (383.23+3pt,117.08-3pt);



\draw (343.32-3pt,116.46-3pt) -- (343.32+3pt,116.46+3pt);
\draw (343.32-3pt,116.46+3pt) -- (343.32+3pt,116.46-3pt);

\draw (296.61-3pt,70.46-3pt) -- (296.61+3pt,70.46+3pt);
\draw (296.61-3pt,70.46+3pt) -- (296.61+3pt,70.46-3pt);

\draw (296.61-3pt,116.46-3pt) -- (296.61+3pt,116.46+3pt);
\draw (296.61-3pt,116.46+3pt) -- (296.61+3pt,116.46-3pt);


\draw (212.9-3pt,71.42-3pt) -- (212.9+3pt,71.42+3pt);
\draw (212.9-3pt,71.42+3pt) -- (212.9+3pt,71.42-3pt);

\draw (259.6-3pt,117.42-3pt) -- (259.6+3pt,117.42+3pt);
\draw (259.6-3pt,117.42+3pt) -- (259.6+3pt,117.42-3pt);

\draw (213.11-3pt,117.29-3pt) -- (213.11+3pt,117.29+3pt);
\draw (213.11-3pt,117.29+3pt) -- (213.11+3pt,117.29-3pt);

\draw (259.39-3pt,71.54-3pt) -- (259.39+3pt,71.54+3pt);
\draw (259.39-3pt,71.54+3pt) -- (259.39+3pt,71.54-3pt);

\draw (236.29-3pt,126.13-3pt) -- (236.29+3pt,126.13+3pt);
\draw (236.29-3pt,126.13+3pt) -- (236.29+3pt,126.13-3pt);

\draw (236.25-3pt,62.12-3pt) -- (236.25+3pt,62.12+3pt);
\draw (236.25-3pt,62.12+3pt) -- (236.25+3pt,62.12-3pt);

\draw (259.49-3pt,37.49-3pt) -- (259.49+3pt,37.49+3pt);
\draw (259.49-3pt,37.49+3pt) -- (259.49+3pt,37.49-3pt);

\draw (213.01-3pt,37.37-3pt) -- (213.01+3pt,37.37+3pt);
\draw (213.01-3pt,37.37+3pt) -- (213.01+3pt,37.37-3pt);

\draw (319.77-3pt,126.1-3pt) -- (319.77+3pt,126.1+3pt);
\draw (319.77-3pt,126.1+3pt) -- (319.77+3pt,126.1-3pt);
\node at (225.27, 30.54) {$h$};

\end{tikzpicture}

    \caption{Leading diagrams contributing to the VEV of the operators $|H|^2$ and $|H|^6$.}
    \label{fig:H2 and H6}
\end{figure}
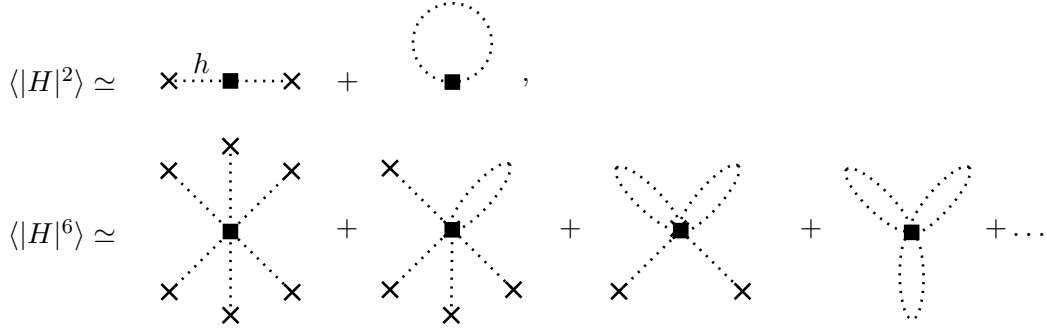

The first category of operators that we study are those whose VEV does not break any approximate symmetry of the SM. Consider for instance $|H|^2$ in the SM Lagrangian or $|H|^6$ in the SMEFT. 
We show the leading contributions to their VEV in Fig.~\ref{fig:H2 and H6}. Those diagrams give approximately\footnote{When writing $\Lambda_H$ we do not distinguish between mass scales and VEVs, because the $O(1)$ couplings in the SM make them numerically comparable.}
\be
\langle |H|^2 \rangle &= & v^2 + c_2\frac{\Lambda_H^2}{16\pi^2}\, , \nn \\
\langle |H|^6 \rangle &= & v^6 + a_6\frac{v^4 \Lambda_H^2}{16\pi^2} +b_6\frac{v^2 \Lambda_H^4}{(16\pi^2)^2} + c_{6}\frac{\Lambda_H^6}{(16\pi^2)^3}  + O\left(\frac{\lambda}{16\pi^2}\right)\, , \label{eq:ex1} 
\ee
where $c_2, a_6, b_6$ and $c_6$ are $O(1)$ numbers. If we include logarithms explicitly Eq.~\eqref{eq:ex1} becomes
\begin{align}
\langle |H|^6 \rangle =&~  v^6 + a_6\frac{v^4 \Lambda_H^2}{16\pi^2}\log^{p_a}\left(\frac{\Lh}{m_h}\right)+b_6\frac{v^2 \Lambda_H^4}{(16\pi^2)^2}\log^{p_b}\left(\frac{\Lh}{m_h}\right)\nonumber\\ 
+&~ c_{6}\frac{\Lambda_H^6}{(16\pi^2)^3}\log^{p_c}\left(\frac{\Lh}{m_h}\right)  + O\left(\frac{\lambda}{16\pi^2}\right)\,,
\label{eq:ex1Logs}
\end{align}
where $p_a\le1$, $p_b\le2$ and $p_c\le3$. Then comparing the leading terms at large $\Lambda_H$ in Eq.~\eqref{eq:ex1Logs}, we obtain that
\begin{equation}
   \Lh\lesssim \frac{4\pi v}{\sqrt{\log\left(\frac{\Lh}{m_h}\right)}}\, ,
\end{equation}
is needed for $|H|^6$ to be a trigger.
These operators are good triggers only if the terms proportional to $v$ dominate over those that contain only the cutoff. 
Whenever the terms proportional to the cutoff in Eq.~\eqref{eq:ex1} dominate, we are in the limit described by Eq.~\eqref{eq:cutoff}, i.e. the VEV of our operator is insensitive to $m_h^2$. We find a sensitivity to the Higgs VEV only if $\Lambda_H \lesssim 4 \pi v$, so these operators can only solve the little hierarchy problem. We are going to encounter many examples of this kind, where an operator VEV is UV-sensitive and does not break any approximate SM symmetry. 
From here on we are not going to include logs in our estimates and just show upper bounds of the type $\Lambda_H \lesssim 4 \pi v$. Being more careful about logs further lowers the upper bounds on $\Lambda_H$ that we list below, since the most superficially divergent diagrams come with the largest powers of logs. The generic situation in our calculations is that contributions to the VEV of an operator proportional to the largest power of the cutoff are those with the largest number of loops and therefore the largest power of $\log \Lambda_H$. Therefore logs further lower our upper bounds on the cutoff and our $\Lambda_H \lesssim 4\pi v$'s are in this sense conservative bounds. This choice does not affect our conclusions, since ultimately we want to show that SMEFT operators are not good triggers.

Before moving on, it is useful to be more explicit about the cutoff $\Lambda_H$. In this work $\Lambda_H$ is the scale where the VEV of our operators and the Higgs mass become calculable. We always take it parametrically smaller than $M_{\rm UV}$, the scale suppressing the $d> 4$ operators in the SMEFT. For definiteness, we imagine that the SMEFT operators are generated at a larger scale $M_{\rm UV}$, so that our calculations are well-defined. In the interest of describing also a simpler and perhaps more realistic UV model, we will often abandon well-defined calculations in favor of estimates where $M_{\rm UV} \simeq \Lambda_{\rm UV}$.

Our prescription for estimating the operators' VEV in the SMEFT immediately generalizes to theories where the UV-sensitive VEVs are calculable. If we are computing the VEV of $|H|^2$ in the low energy theory by adding a $\xi \phi |H|^2$ vertex to the Lagrangian, above $\Lambda_H$ we have to write new couplings that respect the larger symmetry of the theory. For example in a supersymmetric theory we would have the superpotential
\be
W \supset \xi \phi H_u H_d + \mu H_u H_d\, ,
\ee
plus the usual SUSY breaking terms of the MSSM~\cite{Martin:1997ns}. After integrating out both Higgses and Higgsinos at one loop we obtain
\begin{equation}
    V_{\rm EFT}(\xi\phi)=\frac{\mu~(m_{H_u}^2+m_{H_d}^2)}{8\pi^2}~\xi\phi+\ldots
\end{equation}
where $m_{H_{u,d}}^2$ are the soft breaking masses in the Higgs sector~\cite{Martin:1997ns}.
In this theory $\langle |H|^2\rangle$ is calculable, $\Lambda_H$ is well-defined and corresponds to a particular combination of superpotential coefficients and SUSY-breaking parameters. 

A similar reasoning applies to operators containing derivatives, as for example $\mathcal{O}_\square = (H^\dagger H) \square (H^\dagger H)$. The VEV of $\mathcal{O}_\square$ does not break any symmetry and the presence of the derivatives does not change our calculation of the VEV,
\be
\langle \mathcal{O}_\square \rangle = \left(c_{1\square} v^2 + c_{2\square}\frac{\Lambda_H^2}{16\pi^2}\right)\frac{\Lambda_H^4}{16\pi^2} + O\left(\frac{\lambda}{16\pi^2}\right)\, ,
\ee
where the $c_{i\square}$'s are $O(1)$ numbers.
If we want to use this operator as a trigger, the absence of a symmetry leads us again to the result $\Lambda_H \lesssim 4\pi v$. The situation can be different for other operators containing derivatives. For instance we can write $G\widetilde G$ as the divergence of a current. However, what is protecting the $G\widetilde G$ VEV is not the derivative itself, but rather the symmetry associated to the current, as noted in the introduction.

We can now move on to a qualitatively different example. The only dimension 5 operator in the SMEFT breaks lepton number, so if we add it to the SM in isolation and compute its VEV, we obtain $\langle (HL)^2 \rangle = 0$. Lepton number is unbroken in the SM and we did not expect any other result. Similarly, dimension 6 operators that break baryon number~\cite{Weinberg:1979sa} have zero VEV, for instance $\langle QQQL\rangle=0$. If we turn on more than one operator at a time in principle the VEV could be non-zero, but the breaking of the symmetry by $d\geq 4$ operators is hard and we would encounter the problem that we discuss in the next example. Consider
\be
\mathcal{O}_{RR}= (u^{c\dagger} \bar\sigma^\mu c^c)(d^{c\dagger} \bar\sigma_\mu s^c)\, . \label{eq:ORR}
\ee
The VEV of $\mathcal{O}_{RR}$ breaks a subset of the $U(3)^5$ flavor symmetry of the SM kinetic terms and it is proportional to the small parameters that break these symmetries in the SM. However the breaking is hard, which implies that the contributions to the VEV of $\mathcal{O}_{RR}$ that are sensitive to $\langle h \rangle$ and those that are insensitive to it are all proportional to the same small number. The relevant loop diagrams for its VEV are shown in Fig.~\ref{fig:4RFermions} and parametrically they give
\be
\langle \mathcal{O}_{RR} \rangle = C \left(c_{1RR}\frac{\Lambda_H^6}{(16\pi^2)^6}+c_{2RR}\frac{\Lambda_H^4 v^2}{(16\pi^2)^5}+c_{3RR}\frac{\Lambda_H^2 v^4}{(16\pi^2)^4}\right)\, , \label{eq:HardOnly}
\ee
where the $c_{iRR}$'s are $O(1)$ numbers.
The pre-factor $C$ is much smaller than 1, but it is the same for the contributions to $\langle \mathcal{O}_{RR} \rangle$ that are sensitive to $\hv$ and those that are insensitive to it. Therefore there is no parametric limit of the theory where $\Lambda_H$ becomes large and $\mathcal{O}_{RR}$ remains a good trigger (i.e. Eq.~\eqref{eq:TDef} is satisfied). As we smoothly restore the symmetry ($C\to 0$), the upper bound on $\Lambda_H \lesssim 4 \pi v$ implied by Eq.~\eqref{eq:TDef} remains constant, until $\langle \mathcal{O}_{RR} \rangle \lesssim ({\rm meV})^6$ and $\mathcal{O}_{RR}$ can no longer be used to generate any cosmological dynamics that affects the Universe evolution before the present time. Our last example is a potentially good trigger candidate,
\be
\mathcal{O}_{LR}=(Q u^c)(Q d^c)\, ,
\ee
where we take $Q, u^c$ and $d^c$ to have all the same flavor index.
The VEV of $\mathcal{O}_{LR}$ is protected by the chiral symmetry of the quarks, which is broken softly by QCD. There is also a source of hard breaking from the small up and down Yukawa couplings $y_{u,d} \simeq 10^{-5}$. Estimating its VEV gives
\be
\langle (Q u^c)(Q d^c) \rangle = c_{1LR}\Lambda_{\rm QCD}^2 f_\pi^4+ c_{2LR} m_u m_d \frac{\Lambda_H^4}{(16\pi^2)^2}+ c_{3LR} y_u y_d \frac{\Lambda_H^6}{(16\pi^2)^3}\, ,
\label{eq:QuQd}
\ee
where the first contribution is the QCD condensate that breaks the fermion chiral symmetries, the second contribution comes from the diagram in the left panel of Fig.~\ref{fig:O vev} and the third one from the diagram in the right panel of the same Figure. The $c_{iLR}$'s are $O(1)$ numbers. We are including in these coefficients also possible color factors, since when we compute an upper bound on $\Lambda_H$ their effect is suppressed by a fractional power. For example $c_{3LR}\sim N_c^2$, but $\Lambda_H^{\rm max} \propto 1/c_{3LR}^{1/6}\sim 1/N_c^{1/3}.$ Besides, we expect the largest color factors in diagrams with the largest number of loops that are also the most superficially divergent. Therefore color factors typically further lower our upper bounds on the cutoff making our $\Lambda_H \lesssim 4\pi v$ that neglects them, a conservative upper bound.


\input{Fig2}

\begin{figure}[!t]
    \centering

\tikzset{every picture/.style={line width=1pt}} 

\begin{tikzpicture}[x=1pt,y=1pt,yscale=-1,xscale=1]
\tikzset{
  mysq/.style={
    fill=black,
    draw,
    fill opacity=1
  },
  sqsize/.store in=\sqsize,
  sqsize=4.9pt
}

\draw[mysq] (273.25,39.19) rectangle ++(\sqsize,\sqsize);
\draw[mysq] (424.75,39.19) rectangle ++(\sqsize,\sqsize);
\draw    (246.05,23.79) -- (243.85,23.88) ;
\draw [shift={(240.85,24)}, rotate = 357.61] [fill={rgb, 255:red, 0; green, 0; blue, 0 }  ][line width=0.08]  [draw opacity=0] (6.25,-3) -- (0,0) -- (6.25,3) -- cycle    ;
\draw    (241.55,59.24) ;
\draw [shift={(241.55,59.24)}, rotate = 0] [fill={rgb, 255:red, 0; green, 0; blue, 0 }  ][line width=0.08]  [draw opacity=0] (6.25,-3) -- (0,0) -- (6.25,3) -- cycle    ;
\draw    (305.35,23.79) -- (307.1,23.81) ;
\draw [shift={(310.1,23.86)}, rotate = 180.87] [fill={rgb, 255:red, 0; green, 0; blue, 0 }  ][line width=0.08]  [draw opacity=0] (6.25,-3) -- (0,0) -- (6.25,3) -- cycle    ;
\draw    (214.02,37.34) -- (219.3,43.05) ;
\draw    (219.3,37.34) -- (214.02,43.05) ;

\draw   (216.4,41.64) .. controls (216.4,31.78) and (229.67,23.79) .. (246.05,23.79) .. controls (262.43,23.79) and (275.7,31.78) .. (275.7,41.64) .. controls (275.7,51.49) and (262.43,59.49) .. (246.05,59.49) .. controls (229.67,59.49) and (216.4,51.49) .. (216.4,41.64) -- cycle ;
\draw    (310.1,59.2) ;
\draw [shift={(310.1,59.2)}, rotate = 180] [fill={rgb, 255:red, 0; green, 0; blue, 0 }  ][line width=0.08]  [draw opacity=0] (6.25,-3) -- (0,0) -- (6.25,3) -- cycle    ;
\draw    (332.49,38.55) -- (337.77,44.26) ;
\draw    (337.77,38.55) -- (332.49,44.26) ;

\draw   (275.7,41.64) .. controls (275.7,31.78) and (288.97,23.79) .. (305.35,23.79) .. controls (321.73,23.79) and (335,31.78) .. (335,41.64) .. controls (335,51.49) and (321.73,59.49) .. (305.35,59.49) .. controls (288.97,59.49) and (275.7,51.49) .. (275.7,41.64) -- cycle ;

\draw    (417.73,28.76) -- (416.93,28.33) ;
\draw [shift={(414.27,26.94)}, rotate = 27.76] [fill={rgb, 255:red, 0; green, 0; blue, 0 }  ][line width=0.08]  [draw opacity=0] (6.25,-3) -- (0,0) -- (6.25,3) -- cycle    ;
\draw    (434.99,29.34) -- (436.38,28.57) ;
\draw [shift={(439,27.12)}, rotate = 151.03] [fill={rgb, 255:red, 0; green, 0; blue, 0 }  ][line width=0.08]  [draw opacity=0] (6.25,-3) -- (0,0) -- (6.25,3) -- cycle    ;
\draw   (367.9,41.64) .. controls (367.9,31.78) and (381.17,23.79) .. (397.55,23.79) .. controls (413.93,23.79) and (427.2,31.78) .. (427.2,41.64) .. controls (427.2,51.49) and (413.93,59.49) .. (397.55,59.49) .. controls (381.17,59.49) and (367.9,51.49) .. (367.9,41.64) -- cycle ;
\draw    (460.63,59.2) ;
\draw [shift={(460.63,59.2)}, rotate = 180] [fill={rgb, 255:red, 0; green, 0; blue, 0 }  ][line width=0.08]  [draw opacity=0] (6.25,-3) -- (0,0) -- (6.25,3) -- cycle    ;
\draw   (427.2,41.64) .. controls (427.2,31.78) and (440.47,23.79) .. (456.85,23.79) .. controls (473.23,23.79) and (486.5,31.78) .. (486.5,41.64) .. controls (486.5,51.49) and (473.23,59.49) .. (456.85,59.49) .. controls (440.47,59.49) and (427.2,51.49) .. (427.2,41.64) -- cycle ;
\draw  [fill={rgb, 255:red, 0; green, 0; blue, 0 }  ,fill opacity=1 ] (395.9,23.79) .. controls (395.9,22.87) and (396.64,22.14) .. (397.55,22.14) .. controls (398.46,22.14) and (399.2,22.87) .. (399.2,23.79) .. controls (399.2,24.7) and (398.46,25.44) .. (397.55,25.44) .. controls (396.64,25.44) and (395.9,24.7) .. (395.9,23.79) -- cycle ;
\draw  [fill={rgb, 255:red, 0; green, 0; blue, 0 }  ,fill opacity=1 ] (455.2,23.79) .. controls (455.2,22.87) and (455.94,22.14) .. (456.85,22.14) .. controls (457.76,22.14) and (458.5,22.87) .. (458.5,23.79) .. controls (458.5,24.7) and (457.76,25.44) .. (456.85,25.44) .. controls (455.94,25.44) and (455.2,24.7) .. (455.2,23.79) -- cycle ;
\draw    (378.82,27.85) -- (380.21,27.3) ;
\draw [shift={(383,26.21)}, rotate = 158.63] [fill={rgb, 255:red, 0; green, 0; blue, 0 }  ][line width=0.08]  [draw opacity=0] (6.25,-3) -- (0,0) -- (6.25,3) -- cycle    ;
\draw    (475.36,27.8) ;
\draw [shift={(473,26.53)}, rotate = 28.3] [fill={rgb, 255:red, 0; green, 0; blue, 0 }  ][line width=0.08]  [draw opacity=0] (6.25,-3) -- (0,0) -- (6.25,3) -- cycle    ;
\draw    (397.55,59.49) ;
\draw [shift={(394.67,59.6)}, rotate = 357.64] [fill={rgb, 255:red, 0; green, 0; blue, 0 }  ][line width=0.08]  [draw opacity=0] (6.25,-3) -- (0,0) -- (6.25,3) -- cycle    ;
\draw  [draw opacity=0][dash pattern={on 0.84pt off 2.51pt}] (396.93,23.76) .. controls (403.04,17.91) and (414.22,14) .. (427,14) .. controls (439.69,14) and (450.79,17.86) .. (456.93,23.63) -- (427,34) -- cycle ; \draw  [dash pattern={on 0.84pt off 2.51pt}] (396.93,23.76) .. controls (403.04,17.91) and (414.22,14) .. (427,14) .. controls (439.69,14) and (450.79,17.86) .. (456.93,23.63) ;  
\node at (255,18) {$u$};    
\node at (295,17.) {$d$};    
\node at (255,65) {$u^c$};  
\node at (295,64.8) {$d^c$};  
\node at (426.93,20) {$h$};

\end{tikzpicture}

    \caption{Leading diagrams contributing to the VEV of $(Q u^c)(Q d^c)$ if all quarks have the same flavor. The crosses denote quark mass insertions.}
    \label{fig:O vev}
\end{figure}

We find that our operator is a good trigger up to $\Lambda_H \lesssim 4\pi v$ from the comparison between the second and third terms in Eq.~\eqref{eq:QuQd}, the second term being the dominant one that satisfies Eq.~\eqref{eq:TDef}. Even if this attempt did not allow us to raise $\Lambda_H$ by a large factor, it gives us a handle to find useful triggers. There is a limit where $\Lambda_H$ grows arbitrarily large while $\mathcal{O}_{LR}$ remains a good trigger. If we take $y_u\to 0$ then $\mathcal{O}_{LR}$ becomes a good trigger up to $M_{\rm Pl}$. In this limit the chiral symmetry of the up quark is only broken by the QCD condensate which is a monotonic function of the Higgs VEV as long as $\langle h \rangle \gtrsim f_\pi$, as discussed in Appendix~\ref{app:LQCDvsHiggsVEV}, and the VEV of $\mathcal{O}_{LR}$ becomes insensitive to the cutoff $\Lambda_H$. 

In the next Section we follow this strategy and identify the SMEFT triggers with the largest $\Lambda_H$, but, before moving on, it is useful to summarize the examples discussed in this Section. We have seen several qualitatively different possibilities: 

\begin{enumerate}
\item The operator VEV is zero as a consequence of an exact symmetry of the theory. The operator cannot be used as a trigger.
\item The operator VEV is not protected by any symmetry and the operator can only be used to solve the little hierarchy problem $\Lambda_H \lesssim 4\pi v$.
\item The operator VEV is protected by an approximate symmetry and the breaking in the SM is hard. The operator can only be used to solve the little hierarchy problem $\Lambda_H \lesssim 4\pi v$.
\item The operator VEV is protected by an approximate symmetry of the SM. There are two sources of breaking. Hard breaking proportional to small parameters $\epsilon$ and soft breaking proportional to $\hv$. These operators can in principle solve the hierarchy problem up to larger energy scales $\Lambda_H \lesssim 4\pi v/\epsilon$.
\end{enumerate}

It is possible that the breaking of the symmetry is only soft and proportional to $\hv$. This makes the operator a good trigger up to $M_{\rm Pl}$. However, within the SMEFT, these operators have to transform non-trivially under $SU(2)_L$ and they can be used to solve the hierarchy problem only if coupled to new fields that are charged under $SU(2)_L$. This poses the problem of protecting a large hierarchy of scales in the potential of the new states, without making the whole SM supersymmetric or scale invariant. In practice, solutions of the hierarchy problem that resolve this issue have not been proposed and we do not consider this possibility in what follows.

The examples listed above cover most, but not all, cases of interest in the SM and the SMEFT.
In Section~\ref{sec:noT} we consider operators that fail to be triggers for reasons different from the ones discussed in this Section. However, before talking about interesting failures, we describe the vast majority of SMEFT operators up to dimension 6 that fall within the categories described above.

\begin{table}[!t]
\centering
\begin{tabular}{|c|c|c|}
\hline\hline
{\bf Standard Model} & $d \leq 4$ & \\
\hline
No Symmetry & $\Lambda_H$ & \\
\hline
 $|D_\mu H|^2$ & $4\pi v$ & \\
 $|H|^2$ & $4\pi v$ & \\
 $|H|^4$ & $4\pi v$ & \\
 $\Psi^\dagger i \bar\sigma^\mu D_\mu \Psi$ & $4\pi v$ & \\
 ${\rm Tr}[G_{\mu\nu}G^{\mu\nu}]$ & $4\pi v$& \\
 ${\rm Tr}[W_{\mu\nu}W^{\mu\nu}]$ & $4\pi v$  & \\
 $B_{\mu\nu}B^{\mu\nu}$ & $4\pi v$ & \\
 \hline
 Hard Breaking Only & $\Lambda_H$ & Symmetry \\
\hline
 $Q_i H u^c_j$ & $4\pi v$ & Flavor \\
  $Q_i H^\dagger d^c_j$ & $4\pi v$ & Flavor \\
     $L_i H^\dagger e^c_j$ & $4\pi v$ & Flavor \\
  $L H^\dagger e^c$ & $4\pi v$ & Chiral \\
 \hline
 Hard and Soft Breaking & $\Lambda_H$ & Symmetry \\
\hline
 $ G \widetilde G$ & $M_{\rm Pl}$ & CP \\
 $Q H u^c$ & $4\pi v \max\left[1, \left(\frac{\Lambda_{\rm QCD} f_\pi^2}{y_u v^3}\right)^{1/4}\right]$ & Chiral \\
 $Q H^\dagger d^c$ & $4\pi v \max\left[1, \left(\frac{ \Lambda_{\rm QCD} f_\pi^2}{y_d v^3}\right)^{1/4}\right]$ & Chiral \\
\hline\hline
\end{tabular}
\caption{$d\leq 4$ operators in the SM and the maximal cutoff on the Higgs mass $\Lambda_H$ that they can explain as triggers. Whenever flavor indices are specified, they are different ($i \neq j$), when they are not specified, they are the same. Note that the trace in ${\rm Tr}[G_{\mu\nu}G^{\mu\nu}]$ is over $SU(3)$ indices. We do not write it explicitly for $G \widetilde G$, following the definition in Eq.~\eqref{eq:GGdual}.}
\label{table:SM}
\end{table}

    \begin{table}[!t]
\centering
\begin{tabular}{|c|c|}
\hline\hline
{\bf No Symmetry} & $\Lambda_H$\\
\hline
 $(H^\dagger H)^3$ & $4 \pi v$ \\
 $|H^\dagger D_\mu H|^2$ & $4 \pi v$ \\
 $H^\dagger H \square H^\dagger H $ & $4\pi v$ \\
 $H^\dagger H B_{\mu\nu} B^{\mu\nu}$ & $4 \pi v$ \\
 $H^\dagger H {\rm Tr}[W _{\mu\nu} W^{\mu\nu}]$ & $4 \pi v$ \\
 $H^\dagger H {\rm Tr}[G _{\mu\nu} G^{\mu\nu}]$ & $4 \pi v$ \\
 $H^\dagger \tau^I H\, W_{\mu\nu}^I B^{\mu\nu}$ & $4 \pi v$ \\
 $(\psi^\dag\bar\sigma^\mu\psi)^2$ & $4 \pi v$ \\
  $(\psi^\dag\tau^I\bar\sigma^\mu\psi)^2$ & $4 \pi v$ \\
   $(\psi^\dag\bar\sigma^\mu\psi)( \psi^{c\dag} \bar\sigma_\mu \psi^{c})$ & $4 \pi v$ \\
 $( \psi^{c\dag}\bar\sigma^\mu \psi^{c})^2$ & $4 \pi v$ \\
 $H^\dagger\tau^I\,W_{\mu\nu}^I (\psi^c\sigma^{\mu\nu}\psi)$ & $4 \pi v$ \\
$H^\dag B_{\mu\nu}(\psi^c\sigma^{\mu\nu}\psi)$ & $4 \pi v$\\
$(H^\dagger \overleftrightarrow{D}_\mu H) (\psi^\dag\bar\sigma^\mu\psi)$ & $4\pi v$ \\
$(H^\dagger \overleftrightarrow{D}_\mu^I H) (\psi^\dag\tau^I\bar\sigma^\mu\psi)$ & $4\pi v$ \\
$(H^\dagger \overleftrightarrow{D}_\mu H) ( \psi^{c\dag} \bar\sigma^\mu \psi^{c})$ & $4\pi v$ \\
\hline\hline
\end{tabular}
\caption{$d=6$ operators in the SMEFT whose VEV is not protected by any symmetry. These operators can solve the hierarchy problem up to roughly $4\pi v$ (neglecting $O(1)$ factors). The fermions in this table either have all the same flavor or they have the same flavor within a bilinear, i.e. $(u^\dagger \bar \sigma^\mu u)(c^\dagger \bar\sigma_\mu c)$ is in this table.}
\label{table:NoSymmetry}
\end{table}

\begin{table}[!t]
\centering
\begin{tabular}{|c|c| c|}
\hline\hline
{\bf Hard Breaking Only} & $\Lambda_H$ & Symmetry\\
\hline
 $(\psi_i^\dagger \bar\sigma^{\mu}\psi_j)(\psi_k^\dag\bar\sigma_{\mu}\psi_l)$ & $4\pi v$ & Flavor \\
 $(\psi_i^\dag\tau^I\bar\sigma^{\mu}\psi_j)(\psi_k^\dag\tau^I\bar\sigma_{\mu}\psi_l)$ & $4\pi v$ & Flavor \\
  $(\psi_i^\dag\bar\sigma^{\mu}\psi_j)(\psi^{c\dag}_k\bar\sigma_\mu\psi_l^{c})$ & $4\pi v$ & Flavor \\
 $(\psi^{c\dag}_i\bar\sigma^\mu\psi_j^{c})(\psi^{c\dag}_k\bar\sigma_\mu\psi_l^{c})$ & $4\pi v$ & Flavor \\
 $(H^\dagger \overleftrightarrow{D}_\mu H) (\psi^\dag_i\bar\sigma^{\mu}\psi_j)$ & $4\pi v$ & Flavor \\
$(H^\dagger \overleftrightarrow{D}_\mu^I H) (\psi^\dag_i\tau^I\bar\sigma^{\mu}\psi_j)$ & $4\pi v$ & Flavor \\
$(H^\dagger \overleftrightarrow{D}_\mu H) (\psi^c_i\sigma^\mu\psi^{c\dag}_j)$ & $4\pi v$ & Flavor \\
  $(\tilde H^\dagger i D_\mu H)(u_i^{c\dagger} \bar\sigma^\mu d_j^c)$&$ 4\pi v$ & Flavor \\
$(\psi^\dag_i\bar\sigma^{\mu\nu}\psi^{c\dag}_j) \tau^IH\,W_{\mu\nu}^I$ & $4\pi v$ & Flavor \\
 $(\psi^\dag_i\bar\sigma^{\mu\nu}\psi^{c\dag}_j)  H\,B_{\mu\nu}$ & $4\pi v$ & Flavor \\
 $(\psi^\dag_i\bar\sigma^{\mu\nu}T^A\psi^{c\dag}_j)  H\,G_{\mu\nu}^A$ & $4\pi v$ & Flavor \\
 $(L e^c)^\dagger (Q_i d^c_j)$ & $4\pi v$ & Flavor \\
  $(L e^c) (Q_i u^c_j)$ & $4\pi v$ & Flavor \\
  $(H^\dag H)(\psi_i H\psi^c_j)$ & $4\pi v$& Flavor\\
   $(L_i \sigma^{\mu\nu} e^c_j)(Q_k \sigma_{\mu\nu} u^c_l)$ & $4\pi v$ & Flavor \\
 $(Q_i u^c_j)(Q_k d^c_l)$ & $4\pi v$ & Flavor \\
 $(Q_i T^A u^c_j)(Q_k T^A d^c_l)$ & $4\pi v$ & Flavor \\
  $H^\dagger \tau^I H \widetilde W_{\mu\nu}^I B^{\mu\nu}$ & $4 \pi v$ & CP \\
\hline\hline
\end{tabular}
\caption{$d=6$ operators in the SMEFT whose VEV is protected by approximate symmetries of the SM that are broken by dimensionless parameters only. These operators can solve the hierarchy problem up to roughly $4\pi v$ (neglecting $O(1)$ factors). The Latin indices indicate flavor, and at least one pair of indices in the same fermion bilinear must have different values. For example $(u^\dagger \bar \sigma^\mu c)(u^\dagger \bar \sigma^\mu u)$ is in this table, but $(c^\dagger \bar \sigma^\mu c)(u^\dagger \bar \sigma^\mu u)$ is in Table~\ref{table:NoSymmetry}.}
\label{table:HardBreaking}
\end{table}

\begin{table}[!t]
\centering
\begin{tabular}{|c|c|c|}
\hline\hline
{\bf Hard and Soft Breaking} & $\Lambda_H$ & Symmetry \\
\hline
 $(H^\dagger H)(Q H q^c)$ & $4\pi v \max\left[1, \left(\frac{\Lambda_{\rm QCD} f_\pi^2}{y_q v^3}\right)^{1/6}\right]$ & Chiral\\
 $(Q u^c)(Q d^c)$ & $4\pi v \max\left[1, \left(\frac{\Lambda_{\rm QCD} f_\pi^2}{y_u y_d v^3}\right)^{1/6}\right]$ & Chiral \\
 $(Q T^A u^c)(Q T^A d^c)$ & $4\pi v \max\left[1, \left(\frac{\Lambda_{\rm QCD} f_\pi^2}{y_u y_d v^3}\right)^{1/6}\right]$ & Chiral \\
 $(Q^{\dagger}_i \bar\sigma^\mu T^A Q_j)(q^{c\dagger}_i \bar\sigma_\mu T^A q^c_j)$ &$4\pi v\,\text{max}\left[1,\left(\frac{\Lambda_{\rm{QCD}}f_\pi^2}{y_u y_d v^3}\right)^{1/6}\right]$ & Chiral \\
$( \psi^\dag\bar\sigma^{\mu\nu} T^A\,\psi^{c\dag})H\,G_{\mu\nu}^A$ & $4\pi v \,{\rm{max}}\left[1,\left( \frac{m_0^2f_\pi^2\Lambda_{\rm{QCD}} }{y_dg_s v^5}\right)^{1/6}\right]$ & Chiral \\
 $(L \sigma^{\mu\nu} e^c)(Q \sigma_{\mu\nu} u^c)$ & $4\pi v\,\text{max}\left[1,\left(\frac{\Lambda_{\rm{QCD}}f_\pi^2}{y_uv^3}\right)^{1/4}\right]$ & Chiral \\
   $(L e^c)^\dagger (Q d^c)$ & $4\pi v \max\left[1, \left(\frac{ \Lambda_{\rm QCD}f_\pi^2}{y_d v^3}\right)^{1/4}\right]$ & Chiral \\
  $(L e^c) (Q u^c)$ & $4\pi v \max\left[1, \left(\frac{ \Lambda_{\rm QCD}f_\pi^2}{y_u v^3}\right)^{1/4}\right]$ & Chiral \\
\hline\hline
\end{tabular}
\caption{$d=6$ operators in the SMEFT whose VEV is protected by approximate symmetries of the SM that are broken by dimensionless parameters and dimensionful ones proportional to $\langle h \rangle$. These operators can potentially solve the hierarchy problem above $4\pi v$. Lowercase Latin indices indicate flavor. Note that all cutoffs in this table are also $4\pi v$, but we show the full expression to highlight the conceptual differences with respect to the previous Tables.
}
\label{table:HardSoftBreaking}
\end{table}

\begin{table}[!t]
\centering
\begin{tabular}{|c|c|}
\hline\hline
{\bf Unbroken Symmetry} & Symmetry \\
\hline
 $(H L)^2$ & Lepton Number \\
  $QQQL$ & Baryon Number \\
  $d^c u^c u^c e^c$ & Baryon and Lepton Number \\
   $Q Q u^c e^c$ & Baryon and Lepton Number \\
   $Q L u^c d^c$ & Baryon and Lepton Number \\
\hline\hline
\end{tabular}
\caption{$d=5,6$ operators in the SMEFT whose VEV is protected by an unbroken symmetry of the SM.}
\label{table:Unbroken}
\end{table}

\section{The Standard Model up to Dimension Six}\label{sec:SMEFT}

In this Section we show that there are no new\footnote{The old one being $G \widetilde G$.} triggers in the SMEFT up to dimension 6 that can explain the Higgs mass beyond a cutoff $\Lambda_H \simeq 4\pi v$. Our main results are in Tables~\ref{table:SM}, \ref{table:NoSymmetry}, \ref{table:HardBreaking}, \ref{table:HardSoftBreaking} and \ref{table:Unbroken}. They are a straightforward application of the logic outlined in the previous Section, so we will not display all calculations in detail, but just a few examples that extend those already discussed in Section~\ref{sec:examples}. 

It is important to note that we did not compute all $O(1)$ factors in the operators' VEVs, so when we write $4\pi v$ in the tables we really mean $4\pi v/c_{\mathcal{O}}$, where $c_{\mathcal{O}}$ is an $O(1)$ operator-dependent coefficient. We do not find it useful to give all the $O(1)$ coefficients because solving the hierarchy problem with a trigger requires some non-negligible amount of model building and in some cases rather constraining assumptions on the deep UV (i.e. the structure of the landscape of vacua populated in a Multiverse for example).
It seems unlikely that this UV structure would exist solely to explain the difference between $m_h$ and $4\pi v$. However this is our personal point of view and this possibility cannot  be fully ruled-out experimentally.

We begin by examining the SM, i.e. all the operators with $d\leq 4$. We have listed these operators in Table~\ref{table:SM}. Most of them do not break any (approximate) symmetries and can only solve the little hierarchy problem. We have already discussed $|H|^2$ as an example in the previous Section and the other operators in this category are not conceptually different.

In principle also the remaining SM operators fall into one of the categories described in the previous Section, we nevertheless find it useful to discuss explicitly two examples. Consider $\mathcal{O}_F= Q_i H u^c_j$ where $i\neq j$ are flavor indices. We can use (see~\cite{Shifman:1978bx,Shifman:1978by} or \cite{Contino:2010rs, Colangelo:2000dp, deRafael:1997ea} for a review) the general result
\begin{equation}\label{eq:condensate}
    \langle(q \Gamma_r q^c)(q \Gamma_s q^c)\rangle_{\rm QCD}=\frac{\Lambda_{\rm{QCD}}^2f_\pi^4}{(3\times 4)^2}\left[\Tr\Gamma_r\Tr\Gamma_s-\Tr\Gamma_s\Gamma_r\right]\,,
\end{equation}
in the special case $\Gamma_r = \Gamma_s$ to conclude that there is no QCD condensate corresponding to this operator, $\langle \mathcal{O}_F \rangle_{\rm QCD}=0$. Here $\Gamma_{r,s}$ stands for matrices acting on the color, flavor and spinor indices of the quark fields. Eq.~\eqref{eq:condensate} is more than what is needed in our simple $d=4$ case, but it will be useful later when we consider $d=6$ operators.

The fact that the condensate $\langle Q_i u^c_j\rangle$ is vanishing for configurations that break flavor symmetry is not surprising. It follows directly from the Vafa-Witten theorem on vacuum alignment~\cite{Vafa:1983tf}. In vector-like gauge theories, vector symmetries, i.e. $SU(N_f)_V$, are not spontaneously broken. Consequently, any condensate that is non-diagonal in flavor space is forbidden. This leaves just hard breaking of the flavor symmetry from CKM matrix elements. The terms proportional only to the cutoff arise from the first diagram in Fig.~\ref{fig:QHu different flavor}. They are proportional to the same flavor-breaking spurions as all other contributions to the VEV, in particular the second diagram in Fig.~\ref{fig:QHu different flavor} proportional to $\langle h \rangle^2$. In the end this gives $\Lambda_H \lesssim 4\pi v$ for $\mathcal{O}_F$, following the same arguments given for $\mathcal{O}_{RR}$ in the previous Section.

\begin{figure}[t!]
    \centering

\tikzset{every picture/.style={line width=1pt}} 

\begin{tikzpicture}[x=0.85pt,y=0.85pt,yscale=-1,xscale=1]

\draw  [draw opacity=0] (280.31,14.33) .. controls (285.74,12.35) and (292.19,11.2) .. (299.11,11.2) .. controls (317.94,11.2) and (333.3,19.7) .. (334.08,30.35) -- (299.11,31.2) -- cycle ; \draw   (280.31,14.33) .. controls (285.74,12.35) and (292.19,11.2) .. (299.11,11.2) .. controls (317.94,11.2) and (333.3,19.7) .. (334.08,30.35) ;  
\draw  [fill={rgb, 255:red, 0; green, 0; blue, 0 }  ,fill opacity=1 ] (331.66,28.75) -- (336.56,28.75) -- (336.56,33.65) -- (331.66,33.65) -- cycle ;
\draw  [draw opacity=0] (333.99,32.86) .. controls (332.51,43.13) and (317.46,51.2) .. (299.11,51.2) .. controls (292.24,51.2) and (285.83,50.07) .. (280.43,48.12) -- (299.11,31.2) -- cycle ; \draw   (333.99,32.86) .. controls (332.51,43.13) and (317.46,51.2) .. (299.11,51.2) .. controls (292.24,51.2) and (285.83,50.07) .. (280.43,48.12) ;  
\draw []   (280.79,14.12) -- (280.52,49.06) ;
\draw [shift={(280.63,34.79)}, rotate = 270.43] [fill=black ][line width=0.08]  [draw opacity=0] (6.25,-3) -- (0,0) -- (6.25,3) -- cycle    ;
\draw    (311.43,12.41) ;
\draw [shift={(312.41,12.62)}, rotate = 192.2] [fill={rgb, 255:red, 0; green, 0; blue, 0 }  ][line width=0.08]  [draw opacity=0] (6.25,-3) -- (0,0) -- (6.25,3) -- cycle    ;
\draw  [fill={rgb, 255:red, 0; green, 0; blue, 0 }  ,fill opacity=1 ] (279.14,14.12) .. controls (279.14,13.21) and (279.88,12.47) .. (280.79,12.47) .. controls (281.7,12.47) and (282.44,13.21) .. (282.44,14.12) .. controls (282.44,15.03) and (281.7,15.77) .. (280.79,15.77) .. controls (279.88,15.77) and (279.14,15.03) .. (279.14,14.12) -- cycle ;
\draw  [fill={rgb, 255:red, 0; green, 0; blue, 0 }  ,fill opacity=1 ] (278.78,48.12) .. controls (278.78,47.21) and (279.51,46.47) .. (280.43,46.47) .. controls (281.34,46.47) and (282.08,47.21) .. (282.08,48.12) .. controls (282.08,49.03) and (281.34,49.77) .. (280.43,49.77) .. controls (279.51,49.77) and (278.78,49.03) .. (278.78,48.12) -- cycle ;
\draw  [dash pattern={on 0.84pt off 2.51pt}]  (299.19,11.57) .. controls (305.69,30.32) and (317.44,34.57) .. (334.11,31.2) ;
\draw    (285.82,12.72) ;
\draw [shift={(285.82,12.72)}, rotate = 346.56] [fill={rgb, 255:red, 0; green, 0; blue, 0 }  ][line width=0.08]  [draw opacity=0] (6.25,-3) -- (0,0) -- (6.25,3) -- cycle    ;
\draw    (301.91,51.08) ;
\draw [shift={(302.97,51.08)}, rotate = 180] [fill={rgb, 255:red, 0; green, 0; blue, 0 }  ][line width=0.08]  [draw opacity=0] (6.25,-3) -- (0,0) -- (6.25,3) -- cycle    ;
\draw    (279.25,47.91) .. controls (276.82,48.61) and (275.22,47.81) .. (274.45,45.51) .. controls (274.01,43.28) and (272.64,42.38) .. (270.34,42.8) .. controls (268.03,43.04) and (266.88,41.97) .. (266.89,39.6) .. controls (267.24,37.29) and (266.34,35.85) .. (264.19,35.29) .. controls (262.1,34) and (261.84,32.37) .. (263.39,30.41) .. controls (265.26,29.22) and (265.68,27.62) .. (264.64,25.63) .. controls (264.09,23.23) and (265.04,21.81) .. (267.47,21.38) .. controls (269.72,21.46) and (270.94,20.33) .. (271.13,18) .. controls (271.53,15.71) and (272.94,14.86) .. (275.36,15.43) -- (279.14,14.12) ;
\draw  [fill={rgb, 255:red, 0; green, 0; blue, 0 }  ,fill opacity=1 ] (297.54,11.57) .. controls (297.54,10.66) and (298.28,9.92) .. (299.19,9.92) .. controls (300.1,9.92) and (300.84,10.66) .. (300.84,11.57) .. controls (300.84,12.48) and (300.1,13.22) .. (299.19,13.22) .. controls (298.28,13.22) and (297.54,12.48) .. (297.54,11.57) -- cycle ;

\draw  [draw opacity=0] (381.31,14.29) .. controls (386.74,12.31) and (393.19,11.16) .. (400.11,11.16) .. controls (418.94,11.16) and (434.3,19.66) .. (435.08,30.32) -- (400.11,31.16) -- cycle ; \draw   (381.31,14.29) .. controls (386.74,12.31) and (393.19,11.16) .. (400.11,11.16) .. controls (418.94,11.16) and (434.3,19.66) .. (435.08,30.32) ;  
\draw  [fill={rgb, 255:red, 0; green, 0; blue, 0 }  ,fill opacity=1 ] (432.66,28.71) -- (437.56,28.71) -- (437.56,33.61) -- (432.66,33.61) -- cycle ;
\draw  [draw opacity=0] (434.99,32.82) .. controls (433.51,43.09) and (418.46,51.16) .. (400.11,51.16) .. controls (393.24,51.16) and (386.83,50.03) .. (381.43,48.08) -- (400.11,31.16) -- cycle ; \draw   (434.99,32.82) .. controls (433.51,43.09) and (418.46,51.16) .. (400.11,51.16) .. controls (393.24,51.16) and (386.83,50.03) .. (381.43,48.08) ;  
\draw [ ]   (381.79,14.08) -- (381.52,49.03) ;
\draw [shift={(381.63,34.75)}, rotate = 270.43] [fill=black][line width=0.08]  [draw opacity=0] (6.25,-3) -- (0,0) -- (6.25,3) -- cycle    ;
\draw    (412.43,12.37) ;
\draw [shift={(413.41,12.58)}, rotate = 192.2] [fill={rgb, 255:red, 0; green, 0; blue, 0 }  ][line width=0.08]  [draw opacity=0] (6.25,-3) -- (0,0) -- (6.25,3) -- cycle    ;
\draw  [fill={rgb, 255:red, 0; green, 0; blue, 0 }  ,fill opacity=1 ] (380.14,14.08) .. controls (380.14,13.17) and (380.88,12.43) .. (381.79,12.43) .. controls (382.7,12.43) and (383.44,13.17) .. (383.44,14.08) .. controls (383.44,14.99) and (382.7,15.73) .. (381.79,15.73) .. controls (380.88,15.73) and (380.14,14.99) .. (380.14,14.08) -- cycle ;
\draw  [fill={rgb, 255:red, 0; green, 0; blue, 0 }  ,fill opacity=1 ] (379.78,48.08) .. controls (379.78,47.17) and (380.51,46.43) .. (381.43,46.43) .. controls (382.34,46.43) and (383.08,47.17) .. (383.08,48.08) .. controls (383.08,48.99) and (382.34,49.73) .. (381.43,49.73) .. controls (380.51,49.73) and (379.78,48.99) .. (379.78,48.08) -- cycle ;
\draw    (386.82,12.68) ;
\draw [shift={(386.82,12.68)}, rotate = 346.56] [fill={rgb, 255:red, 0; green, 0; blue, 0 }  ][line width=0.08]  [draw opacity=0] (6.25,-3) -- (0,0) -- (6.25,3) -- cycle    ;
\draw    (402.91,51.04) ;
\draw [shift={(403.97,51.04)}, rotate = 180] [fill={rgb, 255:red, 0; green, 0; blue, 0 }  ][line width=0.08]  [draw opacity=0] (6.25,-3) -- (0,0) -- (6.25,3) -- cycle    ;
\draw    (380.25,47.87) .. controls (377.82,48.57) and (376.22,47.77) .. (375.45,45.47) .. controls (375.01,43.25) and (373.64,42.35) .. (371.34,42.77) .. controls (369.03,43) and (367.88,41.94) .. (367.89,39.57) .. controls (368.24,37.25) and (367.34,35.81) .. (365.19,35.25) .. controls (363.1,33.96) and (362.84,32.34) .. (364.39,30.38) .. controls (366.26,29.18) and (366.68,27.58) .. (365.64,25.59) .. controls (365.09,23.19) and (366.04,21.77) .. (368.47,21.34) .. controls (370.72,21.42) and (371.94,20.3) .. (372.13,17.97) .. controls (372.53,15.68) and (373.94,14.82) .. (376.36,15.4) -- (380.14,14.08) ;

\draw (400.19-2.64, 11.54-2.855) -- (400.19+2.64, 11.54+2.855);
\draw (400.19-2.64, 11.54+2.855) -- (400.19+2.64, 11.54-2.855);
\draw [dash pattern={on 0.84pt off 2.51pt}]
      (410.61,31.16) -- (435.11,31.16);
\draw (410.61-2.64, 31.16-2.855) -- (410.61+2.64, 31.16+2.855);
\draw (410.61-2.64, 31.16+2.855) -- (410.61+2.64, 31.16-2.855);

\node[font=\normalsize] at (331.66, 12) {$u_i$};       
\node[font=\normalsize] at (331.66, 50.5) {$u^c_j$};    
\node[font=\normalsize] at (310, 35) {$h$};  
\node[font=\normalsize] at (290, 34) {$d_j$}; 
\node[font=\normalsize] at (255, 25) {$W$};  

\end{tikzpicture}

    \caption{Leading diagrams contributing to the VEV of $Q_i H u_j^c$ if all quarks have different flavor. The crosses denote quark mass insertions.}
    \label{fig:QHu different flavor}
\end{figure}
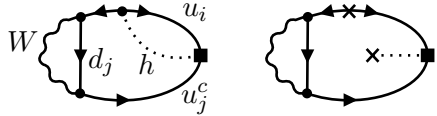

\begin{figure}[!t]
    \centering

\tikzset{every picture/.style={line width=1pt}} 

\begin{tikzpicture}[x=0.75pt,y=0.75pt,yscale=-1,xscale=1]

\tikzset{
  mysq/.style={
    fill=black,
    draw,
    fill opacity=1
  },
  sqsize/.store in=\sqsize,
  sqsize=4.91pt
}
\draw[mysq] (330.8,37.38) rectangle ++(\sqsize,\sqsize);
\draw[mysq] (425.8,37.37) rectangle ++(\sqsize,\sqsize);

\draw[dash pattern={on 0.84pt off 2.51pt}]
      (398.21,40.83) -- (430.71,40.83);
\draw (398.21-2.64, 40.83-2.855) -- (398.21+2.64, 40.83+2.855);
\draw (398.21-2.64, 40.83+2.855) -- (398.21+2.64, 40.83-2.855);
\draw   (268.55,40.83) .. controls (268.55,22.69) and (283.26,7.98) .. (301.4,7.98) .. controls (319.54,7.98) and (334.25,22.69) .. (334.25,40.83) .. controls (334.25,58.97) and (319.54,73.68) .. (301.4,73.68) .. controls (283.26,73.68) and (268.55,58.97) .. (268.55,40.83) -- cycle ;
\draw  [dash pattern={on 0.84pt off 2.51pt}]  (268.55,40.83) -- (334.55,40.83) ;
\draw    (297.17,8.08) ;
\draw [shift={(297.17,8.08)}, rotate = 0] [fill={rgb, 255:red, 0; green, 0; blue, 0 }  ][line width=0.08]  [draw opacity=0] (7.14,-3.43) -- (0,0) -- (7.14,3.43) -- cycle    ;
\draw    (297.2,73.37) ;
\draw [shift={(296.1,73.35)}, rotate = 1] [fill={rgb, 255:red, 0; green, 0; blue, 0 }  ][line width=0.08]  [draw opacity=0] (7.14,-3.43) -- (0,0) -- (7.14,3.43) -- cycle    ;
\draw   (363.55,40.82) .. controls (363.55,22.67) and (378.26,7.97) .. (396.4,7.97) .. controls (414.54,7.97) and (429.25,22.67) .. (429.25,40.82) .. controls (429.25,58.96) and (414.54,73.67) .. (396.4,73.67) .. controls (378.26,73.67) and (363.55,58.96) .. (363.55,40.82) -- cycle ;
\draw    (392.17,8.07) ;
\draw [shift={(392.17,8.07)}, rotate = 0] [fill={rgb, 255:red, 0; green, 0; blue, 0 }  ][line width=0.08]  [draw opacity=0] (7.14,-3.43) -- (0,0) -- (7.14,3.43) -- cycle    ;
\draw    (392.2,73.35) ;
\draw [shift={(391.1,73.33)}, rotate = 1] [fill={rgb, 255:red, 0; green, 0; blue, 0 }  ][line width=0.08]  [draw opacity=0] (7.14,-3.43) -- (0,0) -- (7.14,3.43) -- cycle    ;

\draw  [color={rgb, 255:red, 0; green, 0; blue, 0 }  ,draw opacity=1 ][fill={rgb, 255:red, 0; green, 0; blue, 0 }  ,fill opacity=1 ] (266.32,40.83) .. controls (266.32,39.6) and (267.32,38.6) .. (268.55,38.6) .. controls (269.78,38.6) and (270.78,39.6) .. (270.78,40.83) .. controls (270.78,42.06) and (269.78,43.06) .. (268.55,43.06) .. controls (267.32,43.06) and (266.32,42.06) .. (266.32,40.83) -- cycle ;

\draw (360.905, 37.965) -- (366.195, 43.675);
\draw (366.195, 37.965) -- (360.905, 43.675);

\node at (330.8, 12) {$u$};     
\node at (330.8, 68) {$u^c$};   
\node at (301.4, 32) {$h$};

\end{tikzpicture}

    \caption{Leading diagrams contributing to the VEV of $Q H u^c$ if all quarks have the same flavor. The crosses denote quark mass insertions.}
    \label{fig:QHu same flavor}
\end{figure}
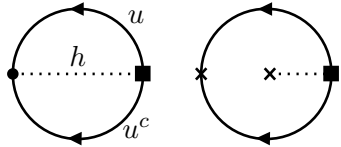

A close relative of the previous operator, $\mathcal{O}_Y = Q H u^c$, with $Q$ and $u^c$ sharing the same flavor index, has a VEV that breaks the chiral symmetry of the up quark, but preserves other flavor symmetries. The QCD condensate breaks this symmetry and we find
\be
\langle Q H u^c \rangle \simeq \Lambda_{\rm QCD}^2 v f_\pi +\ldots\, .
\ee
The leading term proportional to the cutoff (for $\Lambda_H \gg v$) is
\be
\langle Q H u^c \rangle \simeq c_{1Y} y_u\frac{\Lambda_H^4}{(16 \pi^2)^2}+\ldots\, . \label{eq:YUV}
\ee
Putting these two contributions to the VEV together and setting $c_{1Y}=1$ we obtain
\be
\Lambda_H \lesssim 4\pi v \left(\frac{\Lambda_{\rm QCD} f_\pi^2}{y_u v^3}\right)^{1/4} \simeq v\, .\label{eq:LHY}
\ee
This does not mean that the cutoff is smaller than $4\pi v$, but rather that the soft breaking of the chiral symmetry from QCD is subdominant to loop diagrams proportional to some powers of $v$ and a hard breaking parameter. If we set the Higgs to its VEV in both vertices of the previous loop diagram (see Fig.~\ref{fig:QHu same flavor}) we obtain
\be
\langle Q H u^c \rangle \simeq c_{2Y} y_u\frac{\Lambda_H^2 v^2}{(16 \pi^2)}+\ldots \label{eq:YUV2}
\ee
which gives $\Lambda_H \lesssim 4\pi v$ if compared with Eq.~\eqref{eq:YUV}. The results for the other operators in Table~\ref{table:SM} can be obtained either following one of these two examples or one of those in Section~\ref{sec:examples}. As noted in the introduction, we do not compute the coefficients $c_{iY}$. We do not expect a full calculation to alter our conclusions because to increase the bound in Eq.~\eqref{eq:LHY} by a factor of 10 we would need $c_{1Y}\simeq 10^{-4}$. Similarly, to increase by a factor of 10 the upper bound on $\Lambda_H$ obtained by comparing Eq.~\eqref{eq:YUV} with Eq.~\eqref{eq:YUV2}, we would need a hierarchy $c_{1Y}/c_{2Y} \simeq 10^{-2}$ from two diagrams that are almost identical (see Fig.~\ref{fig:QHu same flavor}). Similar arguments apply to all operators considered in this work.

\begin{figure}[t!]
    \centering

\tikzset{every picture/.style={line width=1pt}} 

\begin{tikzpicture}[x=0.75pt,y=0.75pt,yscale=-1,xscale=1]
\tikzset{
  mysq/.style={
    fill=black,
    draw,
    fill opacity=1
  },
  sqsize/.store in=\sqsize,
  sqsize=4.91pt
}

\draw   (173.33,55.83) .. controls (173.33,37.7) and (188.03,23) .. (206.17,23) .. controls (224.3,23) and (239,37.7) .. (239,55.83) .. controls (239,73.97) and (224.3,88.67) .. (206.17,88.67) .. controls (188.03,88.67) and (173.33,73.97) .. (173.33,55.83) -- cycle ;
\draw[mysq] (234.3,52.13) rectangle ++(\sqsize,\sqsize);
\draw[mysq] (459.82,52.13) rectangle ++(\sqsize,\sqsize);
\draw[mysq] (344.97,52.13) rectangle ++(\sqsize,\sqsize);
\draw  [dash pattern={on 0.84pt off 2.51pt}]  (173.33,55.83) -- (239,55.83) ;
\draw    (182.33,33.05) -- (181.12,34.6) ;
\draw [shift={(179.26,36.96)}, rotate = 308.17] [fill={rgb, 255:red, 0; green, 0; blue, 0 }  ][line width=0.08]  [draw opacity=0] (7.14,-3.43) -- (0,0) -- (7.14,3.43) -- cycle    ;
\draw    (179.19,74.19) ;
\draw [shift={(179.19,74.19)}, rotate = 51.34] [fill={rgb, 255:red, 0; green, 0; blue, 0 }  ][line width=0.08]  [draw opacity=0] (7.14,-3.43) -- (0,0) -- (7.14,3.43) -- cycle    ;
\draw   (398.86,56.41) .. controls (398.86,38.27) and (413.56,23.57) .. (431.69,23.57) .. controls (449.82,23.57) and (464.52,38.27) .. (464.52,56.41) .. controls (464.52,74.54) and (449.82,89.24) .. (431.69,89.24) .. controls (413.56,89.24) and (398.86,74.54) .. (398.86,56.41) -- cycle ;
\draw    (454.43,33.34) -- (453.06,31.78) ;
\draw [shift={(451.07,29.53)}, rotate = 48.59] [fill={rgb, 255:red, 0; green, 0; blue, 0 }  ][line width=0.08]  [draw opacity=0] (7.14,-3.43) -- (0,0) -- (7.14,3.43) -- cycle    ;
\draw    (411.46,82.02) -- (414.31,84.04) ;
\draw [shift={(416.75,85.78)}, rotate = 215.43] [fill={rgb, 255:red, 0; green, 0; blue, 0 }  ][line width=0.08]  [draw opacity=0] (7.14,-3.43) -- (0,0) -- (7.14,3.43) -- cycle    ;
\draw    (415.86,27.62) ;
\draw [shift={(415.86,27.62)}, rotate = 141.34] [fill={rgb, 255:red, 0; green, 0; blue, 0 }  ][line width=0.08]  [draw opacity=0] (7.14,-3.43) -- (0,0) -- (7.14,3.43) -- cycle    ;
\draw    (455.25,79.28) ;
\draw [shift={(453.93,80.87)}, rotate = 309.62] [fill={rgb, 255:red, 0; green, 0; blue, 0 }  ][line width=0.08]  [draw opacity=0] (7.14,-3.43) -- (0,0) -- (7.14,3.43) -- cycle    ;
\draw    (428.86,86.77) -- (434.14,92.48) ;
\draw    (434.14,86.77) -- (428.86,92.48) ;

\draw    (429.37,20.77) -- (434.65,26.48) ;
\draw    (434.65,20.77) -- (429.37,26.48) ;

\draw   (284,55.97) .. controls (284,37.84) and (298.7,23.14) .. (316.83,23.14) .. controls (334.97,23.14) and (349.67,37.84) .. (349.67,55.97) .. controls (349.67,74.11) and (334.97,88.81) .. (316.83,88.81) .. controls (298.7,88.81) and (284,74.11) .. (284,55.97) -- cycle ;
\draw    (293,33.19) -- (291.78,34.74) ;
\draw [shift={(289.93,37.1)}, rotate = 308.17] [fill={rgb, 255:red, 0; green, 0; blue, 0 }  ][line width=0.08]  [draw opacity=0] (7.14,-3.43) -- (0,0) -- (7.14,3.43) -- cycle    ;
\draw    (289.86,74.34) ;
\draw [shift={(289.86,74.34)}, rotate = 51.34] [fill={rgb, 255:red, 0; green, 0; blue, 0 }  ][line width=0.08]  [draw opacity=0] (7.14,-3.43) -- (0,0) -- (7.14,3.43) -- cycle    ;
\draw    (281.37,53.03) -- (286.65,58.74) ;
\draw    (286.65,53.03) -- (281.37,58.74) ;

\draw    (396.7,53) -- (401.99,58.71) ;
\draw    (401.99,53) -- (396.7,58.71) ;

\draw    (464.52,56.41) .. controls (467.05,56.6) and (468.17,57.87) .. (467.89,60.24) .. controls (467.26,62.41) and (468.06,63.94) .. (470.31,64.81) .. controls (472.44,65.88) and (472.91,67.45) .. (471.72,69.52) .. controls (470.31,71.33) and (470.49,72.95) .. (472.26,74.38) .. controls (473.89,76.11) and (473.76,77.79) .. (471.88,79.44) .. controls (469.93,80.48) and (469.47,82.03) .. (470.5,84.1) .. controls (471.15,86.44) and (470.23,87.89) .. (467.74,88.46) .. controls (465.55,88.25) and (464.28,89.23) .. (463.93,91.42) .. controls (462.84,93.65) and (461.27,94.12) .. (459.24,92.85) .. controls (457.56,91.27) and (455.88,91.24) .. (454.2,92.77) .. controls (452.03,94.03) and (450.35,93.61) .. (449.15,91.51) .. controls (448.38,89.44) and (446.79,88.76) .. (444.37,89.49) -- (441.09,87.71) ;
\draw    (349.67,55.97) .. controls (352.2,56.16) and (353.32,57.44) .. (353.03,59.81) .. controls (352.4,61.98) and (353.2,63.51) .. (355.45,64.38) .. controls (357.59,65.45) and (358.06,67.02) .. (356.86,69.09) .. controls (355.46,70.9) and (355.64,72.52) .. (357.41,73.95) .. controls (359.04,75.68) and (358.91,77.36) .. (357.03,79.01) .. controls (355.08,80.05) and (354.61,81.6) .. (355.64,83.66) .. controls (356.3,86.01) and (355.38,87.46) .. (352.89,88.03) .. controls (350.69,87.81) and (349.42,88.8) .. (349.07,90.99) .. controls (347.98,93.22) and (346.41,93.69) .. (344.38,92.42) .. controls (342.7,90.84) and (341.02,90.81) .. (339.35,92.34) .. controls (337.18,93.6) and (335.49,93.18) .. (334.3,91.08) .. controls (333.53,89.01) and (331.93,88.33) .. (329.52,89.06) -- (326.23,87.28) ;
\draw    (239,55.83) .. controls (241.53,56.02) and (242.65,57.3) .. (242.36,59.66) .. controls (241.73,61.83) and (242.54,63.36) .. (244.78,64.24) .. controls (246.92,65.31) and (247.39,66.88) .. (246.19,68.95) .. controls (244.79,70.76) and (244.97,72.38) .. (246.74,73.81) .. controls (248.37,75.53) and (248.24,77.22) .. (246.36,78.87) .. controls (244.41,79.91) and (243.95,81.46) .. (244.98,83.52) .. controls (245.63,85.87) and (244.71,87.32) .. (242.22,87.89) .. controls (240.02,87.67) and (238.75,88.66) .. (238.41,90.85) .. controls (237.31,93.08) and (235.74,93.55) .. (233.71,92.28) .. controls (232.03,90.7) and (230.35,90.67) .. (228.68,92.2) .. controls (226.51,93.46) and (224.82,93.04) .. (223.63,90.94) .. controls (222.86,88.87) and (221.26,88.19) .. (218.85,88.92) -- (215.57,87.14) ;
\draw  [dash pattern={on 0.84pt off 2.51pt}]  (432.44,55.83) -- (462.275, 54.83) ;
\draw (432.44-2.64, 55.83-2.855) -- (432.44+2.64, 55.83+2.855);
\draw (432.44-2.64, 55.83+2.855) -- (432.44+2.64, 55.83-2.855);
\draw[dash pattern={on 0.84pt off 2.51pt}] (317.59,55.83) -- (347.425,55.83);
\draw (317.59-2.64, 55.83-2.855) -- (317.59+2.64, 55.83+2.855);
\draw (317.59-2.64, 55.83+2.855) -- (317.59+2.64, 55.83-2.855);

\node at (195.4, 48.) {$h$};  
\draw (215.33,32.54) node [anchor=north west][inner sep=0.75pt]    {$d$};
\draw (215.33,65.07) node [anchor=north west][inner sep=0.75pt]    {$d^{c}$};
\draw (245.5,50.4) node [anchor=north west][inner sep=0.75pt]    {$Z$};
\draw[color={rgb,255:red,0; green,0; blue,0}, draw opacity=1,
      fill={rgb,255:red,0; green,0; blue,0}, fill opacity=1]
  (171.10,55.83) .. controls (171.10,54.60) and (172.10,53.60) .. (173.33,53.60)
  .. controls (174.56,53.60) and (175.56,54.60) .. (175.56,55.83)
  .. controls (175.56,57.06) and (174.56,58.06) .. (173.33,58.06)
  .. controls (172.10,58.06) and (171.10,57.06) .. (171.10,55.83)
  -- cycle ;

\end{tikzpicture}
    \caption{Diagrams contributing to the VEV of $\mO_{QHW}$ in Eq.~\eqref{eq:O2}. The first diagram gives a contribution of order $\Lh^6$, while the following ones have more powers of $v$.}
    \label{fig:O2}
\end{figure}
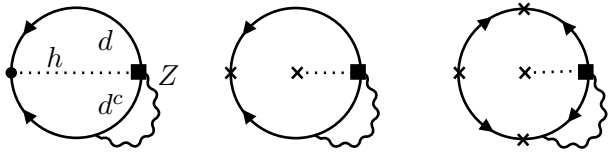

We can move on to Table~\ref{table:NoSymmetry}. In the Table, all results follow the same logic that we have already outlined for $|H|^2$ and $|H|^6$ in Section~\ref{sec:examples}. Take for instance
\be
\mO_{QHW}=(Q^\dag\bar\sigma^{\mu\nu}d^{c\dag})\tau^I H W_{\mu\nu}^I\,. 
\ee
Its VEV does not break any symmetry and we expect contributions proportional to the cutoff and to the Higgs VEV, indeed we find
\begin{equation}\label{eq:O2}
\langle\mO_{QHW}\rangle=\frac{g^2 y_d}{\sqrt{g^2+g'^2}}\left( c_{QHW}^{(1)}\frac{\Lh^6}{(16\pi^2)^3}+c_{QHW}^{(2)} v^2\frac{\Lh^4}{\Loop^2}+c_{QHW}^{(3)}v^4y_d^2\frac{\Lh^2}{\Loop^2}\right)+\dots\,,
\end{equation}
from the diagrams in Fig.~\ref{fig:O2}, where the ellipsis corresponds to terms that are independent of the cutoff $\Lh$. Therefore, from the comparison between the first and the second terms in Eq.~\eqref{eq:O2}, we find $\Lh\lesssim 4\pi v$. Also in this case we did not compute the coefficients $c_{QHW}^{(i)}$, but altering our upper bound requires a large hierarchy between coefficients coming from very similar diagrams. This argument applies to all operators in Tables~\ref{table:NoSymmetry} and~\ref{table:HardBreaking}.

In Table~\ref{table:HardBreaking} we show operators whose VEV transforms non-trivially under symmetries which are broken only by dimensionless parameters within the SM. We have already seen an example in Eq.~\eqref{eq:HardOnly}. We can complete the example by noting that this operator does not get a VEV from QCD, as we can see from Eq.~\eqref{eq:condensate}, where $\Gamma_{r}^{ij}=\Gamma_s^{ij}=\delta^{i1}\delta^{j2}$. The VEV is still generated perturbatively, but all contributions are proportional to the same flavor-breaking spurions, as shown in Eq.~\eqref{eq:HardOnly}.
In the end we still get $\Lambda_H \lesssim 4\pi v$.

To illustrate Table~\ref{table:HardSoftBreaking} we choose an example slightly different from before. We start from 
\begin{equation}
    \mO_{LQ}=(L\sigma^{\mu\nu}e^c)(Q\sigma_{\mu\nu}u^c) \, .
\end{equation}
From standard identities for two-component Weyl spinors it follows that
\begin{align}
    \mO_{LQ}&=-2(Lu^c)(Qe^c)-(Le^c)(Qu^c)\, ,
    \label{eq:O7}
\end{align}
where the parentheses show Lorentz contractions.
Both operators get a VEV, but our interest lies in the second operator. The breaking of the chiral symmetry of the quarks is both hard (from the corresponding Yukawa) and soft (from the QCD condensate) within the SM. So we are going to have a contribution to the VEV proportional to $y_u \Lambda_H^3$ and one to $\Lambda_{\rm QCD} f_\pi^2$ that differ parametrically in the spurions breaking the symmetry. This is identical to the operator $(Q u^c)(Q d^c)$ already discussed in Section~\ref{sec:examples}. The difference lies in the first lepton bilinear. The breaking of the chiral symmetry on the leptons is only hard, so both contributions to the VEV of $\mO_{LQ}$ are proportional to the cutoff. Overall we have 
\be
\langle \mO_{LQ}\rangle = \frac{  \Lambda_H^2}{16\pi^2}\left(c_{LQ}^{(1)}\frac{y_u y_e \Lambda_H^4}{(16\pi^2)^2}+c_{LQ}^{(2)}\frac{m_u m_e \Lambda_H^2}{16\pi^2}+c_{LQ}^{(3)} m_e \Lambda_{\rm QCD}f_\pi^2\right)\, ,
\ee
from diagrams with the same topology as those in Fig.~\ref{fig:O vev}. Note that $c_{LQ}^{(1,2)}$ get contributions from both operators in Eq.~\eqref{eq:O7}. We can see immediately the hard+soft breaking of the quark chiral symmetry is enough to place this operator in Table~\ref{table:HardSoftBreaking}, i.e. there exists a limit ($y_u \to 0$) where this operator is a good trigger up to $M_{\rm Pl}$. Unfortunately within the SM the size of $y_u$ is not sufficiently small to raise the cutoff above $4\pi v$,
\begin{equation}
    \Lh \lesssim 4\pi v\,\text{max}\left[1,\left(\frac{\Lambda_{\rm{QCD}}f_\pi^2}{y_uv^3}\right)^{1/4}\right] = 4\pi v\,.
\end{equation}
Note that this is true for all the operators in Table~\ref{table:HardSoftBreaking}. Unfortunately the size of the SM parameters leaves all cutoffs at $4\pi v$. We chose to write the full expression in the Table to highlight the parametric and conceptual differences with respect to Table~\ref{table:HardBreaking}. Once again we do not expect the calculation of $c_{\mathcal{O}_{LQ}}^{(i)}$ and similar calculations for the other operators in Table~\ref{table:HardSoftBreaking} to alter our conclusions, because they enter our estimates as $\Lambda_H^{\rm max}\sim c^{-1/n}$, with $n=4,6$.

The operators in Table~\ref{table:Unbroken} have zero VEV within the SM, since their VEV would break lepton or baryon number. Even if we turn on more than one operator at a time and try to get a VEV through interference effects, this would not raise the cutoff above $4\pi v$, since each one of these operators would behave as a source of hard breaking. The only exception that we could find is the interference between $W\widetilde W$ and $QQQL$ that we discuss in the next Section.

\section{Non-Trivial Failures}\label{sec:noT}

We have briefly discussed how $G\widetilde G$ is the only trigger operator in the SM that can solve the hierarchy problem all the way to $M_{\rm Pl}$. This suggests an  alternative that we have not yet discussed, i.e. the antisymmetric contraction of two electroweak bosons field strengths, $W \widetilde W$. Purely within the SM this operator has zero VEV, because there are no explicit sources of $B+L$ breaking in addition to electroweak sphalerons \cite{FileviezPerez:2014xju}.  

There are several ways to see this more explicitly. The most direct one consists once again in coupling a probe scalar to $W \widetilde W$
\be
\mathcal{L} \supset -\frac{\alpha_W}{8\pi}\left(\xi \phi+\theta_W\right) W \widetilde W\, . \label{eq:WW}
\ee
We can then perform a $U(1)_{B+L}$ rotation that contains $\phi$,
\be
\psi \to e^{i Q_{B+L}^\psi \alpha_\psi(x)}\psi\, , \quad \sum_\psi Q_{B+L}^\psi\alpha_\psi(x)= \xi\phi(x)+\theta_W\, ,
\ee
where $\psi$ are all the SM fermions charged under $B+L$ and $Q_{B+L}^\psi$ are their charges. The anomalous transformation of the path integral measure under this field redefinition removes the $\xi\phi W\widetilde W$ term in the Lagrangian and the constant $\theta_W$ term. $\phi$ remains in the theory, but it is now derivatively coupled to SM fermions through the $B+L$ current,
\be
\partial_\mu \phi \, J^\mu_{B+L}\, .
\ee
If we now integrate out all SM fields, we do not generate any potential for $\phi$, but just derivative couplings.
To remedy this situation, and give $W \widetilde W$ a VEV, we can add to the SM a new source of $B+L$ breaking. The simplest possibility is the dimension 6 operator \cite{Weinberg:1979sa}
\be
c_{BL}\frac{(Q Y_u Q) (Q Y_d L)}{M_{\rm UV}^2}\, ,
\ee
which has the right quantum numbers to interfere with the ’t Hooft vertex of $SU(2)_L$ and generate a non-zero potential for $\phi$ in Eq.~\eqref{eq:WW}. In addition to the $O(1)$ Wilson coefficient $c_{BL}$, we have included the Yukawa matrices $Y_{u,d}$ to have the same normalization for the operator that one obtains when integrating out a color triplet Higgs of mass $M_{\rm UV}$ in a Grand Unified Theory\footnote{To be precise $Y_u Y_d$ are $3\times 3$ flavor matrices that determine the flavor structure of the $B+L$-breaking operator as $(Q_i (Y_u)_{ij} Q_j)(Q_k (Y_d)_{km} L_m)$. In a GUT they are the triplet Higgs $T$ Yukawa couplings, $Y_u QQT+Y_d Q L T$. At the GUT scale they are numerically identical to $Y_{u,d}$ that then run down to become the up and down quarks Yukawa matrices.}. Bounds from proton decay place the triplet around $10^{17}$~GeV~\cite{PhysRevD.105.112012, PhysRevD.102.112011, PhysRevD.95.012004, PhysRevD.96.012003, Murayama:2001ur}.

It is a simple exercise of instanton NDA~\cite{Csaki:2023ziz} or standard instanton calculus~\cite{Affleck:1980mp,Coleman:1985rnk,Espinosa:1989qn,Morrissey:2005uza,Csaki:2019vte,Sesma:2024tcd} to integrate out the SM fields and obtain in the dilute instanton gas approximation
\be
V(\xi \phi) &\simeq& c_{BL}^3\frac{\det Y_u Y_d(M_{\rm UV})}{(4\pi)^6}\left(\frac{2\pi}{\alpha_W(M_{\rm UV})}\right)^4 \cos(\xi\phi(x)+\theta_W) \nn \\&\times&\left[ \left(\frac{\alpha_W(M_{\rm UV})}{\alpha_W(\sqrt{2}\pi v)}\right)^4 \left(\frac{\det Y_u Y_d(\sqrt{2}\pi v)}{\det Y_u Y_d(M_{\rm UV})}\right) v^2 (\sqrt{2}\pi v)^2\left(\frac{\sqrt{2}\pi v}{M_{\rm UV}}\right)^6 e^{- \frac{2\pi}{\alpha_W(\sqrt{2}\pi v)}}\right. \nn \\
&+&\left.M_{\rm UV}^4 e^{- \frac{2\pi}{\alpha_W(M_{\rm UV})}} \right]\, , \label{eq:WWvev}
\ee
where we assumed that the mass scale $M_{\rm UV}$ differs from the corresponding scale with dimension of VEV by an unspecified $O(1)$ coupling.
The factor $\det Y_u Y_d$ is a product over the Yukawas of all generations and can be deduced from the selection rules of the anomalous symmetries of the SM~\cite{Csaki:2024ywk}. We could have absorbed this factor in $M_{\rm UV}$, but that would have hidden the breaking of SM chiral symmetries needed to obtain a non-zero result.

If we assume SM running, the UV term proportional to $M_{\rm UV}^4$ dominates the potential, making it insensitive to the Higgs VEV for any $M_{\rm UV}$ far above the weak scale.
We can make $W\widetilde W$ a trigger if we alter the running of $\alpha_W$ such that\footnote{Here we are neglecting the two-loop running of $\alpha_W$ and the one-loop running of $\det Y_u Y_d$ to give a simple parametric estimate of what is required of $\alpha_W$, but the latter can be numerically important.}
\be
\alpha_W(M_{\rm UV})\lesssim \frac{\alpha_W(\sqrt{2}\pi v)}{1+\frac{\alpha_W(\sqrt{2}\pi v)}{2\pi}\log \frac{M_{\rm UV}^{10}}{(\sqrt{2} \pi)^8 v^{10}}}\, . \label{eq:aW}
\ee
In this limit the IR contribution to $V(\xi \phi)$ dominates. However this requires adding to the SM a large number of particles charged under $SU(2)_L$~\cite{Csaki:2023ziz}. We do not comment on the necessary model-building any further, because even if we alter the running of $\alpha_W$ appropriately, the IR contribution to the VEV of $W \widetilde W$ is too small to affect the evolution of the Universe. 
From Eq.~\eqref{eq:WWvev} we can compute this contribution to the VEV, setting $c_{BL}=1$ we obtain
\be
\left.\langle W \widetilde W\rangle \right|_{\rm IR} &=& \theta_W \frac{\det Y_u Y_d}{(4\pi)^6}\left(\frac{2\pi}{\alpha_W(\sqrt{2}\pi v)}\right)^4 v^2(\sqrt{2}\pi v)^2 \left(\frac{\sqrt{2}\pi v}{M_{\rm UV}}\right)^6 e^{- \frac{2\pi}{\alpha_W(\sqrt{2}\pi v)}} \nn \\
&\simeq & (10^{-34}\;{\rm eV})^4\theta_W \det Y_u Y_d \left(\frac{10^{17}\;{\rm GeV}}{M_{\rm UV}}\right)^6 \left(\frac{v}{174\;{\rm GeV}}\right)^{10}\, . \label{eq:WWvevN}
\ee
Even if we imagine that $(\theta_W \det Y_u Y_d) = O(1)$, this result is way below the value of the CC, so the interaction of $\phi$ with $W\widetilde W$ cannot generate any interesting cosmological dynamics.
The previous calculation shows that $d\langle W \widetilde W\rangle/d\log \langle h \rangle = O(10)$ for any value of $\Lambda_H$, if we let $\alpha_W$ run to zero in the UV sufficiently fast, i.e. we satisfy Eq.~\eqref{eq:aW}.
Therefore, strictly speaking, $W\widetilde W$ is a trigger with $\Lambda_H \simeq M_{\rm Pl}$, in the SM plus the operator $QQQL$ plus enough new states to alter the running of $\alpha_W$. However we do not know how to use $W\widetilde W$ to select the observed $m_h^2$, because the cosmological dynamics generated by its VEV is too slow to affect the evolution of the Universe. We hope that this discussion will inspire our readers to find a creative use of $W \widetilde W$. 
Alternatives to $QQQL$ that break $B+L$ at dimension 6 give smaller results for $\langle W\widetilde W \rangle$ or still do not make $W\widetilde W$ a viable trigger for other reasons (for instance because they do not allow to close all the legs in the 't Hooft vertex). However it is interesting to consider $(QQQL)^3/M_{\rm UV}^{14}$ so that $B+L$ mod 3 is preserved as in the SM and the proton does not decay. In this case we still find an induced potential proportional to $v^{18}$ which is smaller than the CC even for $M_{\rm UV}\simeq$~TeV. Additionally, the large operator dimension makes the overall result for $\langle W \widetilde W\rangle$ UV-dominated and insensitive to $v$, unless SM running is altered even more than in Eq.~\eqref{eq:aW}.

There is a second operator that is not a trigger and belongs in Table~\ref{table:HardSoftBreaking}, but requires more work than those described in the previous Section to be excluded as a trigger. The Weinberg gluonic operator~\cite{WeinbergGluon},
\begin{equation}\label{eq:Weinberg operator}
\mathcal{O}_{\mathcal{W}}(x)=f^{abc}\varepsilon^{\alpha\beta\gamma\delta}G^a_{\mu\alpha}(x)G^b_{\beta\gamma}(x)G_\delta^{\mu,c}(x)\,,
\end{equation}
is CP-odd and the only other CP-odd vertex in QCD is $\theta G\widetilde G$. However, $G\widetilde G$ is a total derivative and does not enter computations at any order in perturbation theory. It gives a contribution to the $\mO_{\mathcal{W}}$ VEV that we estimate using a dilute instanton gas approximation matched to experimental results in the non-perturbative regime. The leading contribution to $\langle \mathcal{O}_{\mathcal{W}} \rangle$ which is sensitive to the cutoff is due to CP violation in weak interactions. In total we find two main contributions to the VEV in the form
\begin{equation}   \langle\mO_{\mathcal{W}}\rangle=\widetilde \chi_{QW}\bar{\theta} m_u m_d \Lambda_{\rm QCD}^2 f_\pi^2 +c_{\mathcal{W}}\frac{\mathcal{J} G_F^2}{1536\pi^6}\frac{m_b^2m_s^2m_d^2}{m_W^4}\frac{g_s^3\Lh^8}{\Loop^2}\,, \label{eq:OW}
\end{equation}
where $\mathcal{J}={\rm Im}\left(V_{us}V_{cb}V_{ub}^*V_{cs}^*\right)$ is the usual Jarlskog invariant, the only CP-odd, flavor invariant combination of the Yukawa matrices, and $V$ is the CKM matrix~\cite{Cabibbo:1963yz}\footnote{Note the difference compared to the renormalization of $\bar \theta$. The Weinberg gluonic operator is not a total derivative and can be generated at three loops. UV sensitive loop contributions to $\bar \theta$ are instead part of the renormalization of ${\rm Arg\; \rm det} Y_u Y_d$, so one has to go through every quark mass, which in the SM requires going to seven loops and introduces additional Yukawa suppressions in the Jarlskog invariant $J=6(y_t^2-y_c^2)(y_t^2-y_u^2)(y_c^2-y_u^2)(y_b^2-y_s^2)(y_b^2-y_d^2)(y_s^2-y_d^2)\mathcal{J}$.}. The dimensionless factors $c_{\mathcal{W}}$ and $\tilde \chi_{QW}$ are given in Appendix~\ref{app:examples} and~\ref{app:Weinberg}, respectively. The perturbative calculation of the $\Lambda_H^8$ term was performed in~\cite{Pospelov:1994uf}. We compute $\widetilde \chi_{QW}$ in Appendix~\ref{app:Weinberg} using a dilute instanton gas approximation with a cutoff on the maximal size of the instantons determined by matching the chiral susceptibility to experimental data. From Eq.~\eqref{eq:OW} we find \be
\Lh&\lesssim &4\pi v\,{\rm{max}}\left[1,\left(\frac{1536\pi^6(16\pi^2)^2\tilde\chi_{QW}\bar\theta m_u m_d{\Lambda_{\rm{QCD}}}^2f_\pi^2}{\mathcal{J} g_s^3 c_\mathcal{W}}\frac{G_F^2m_W^4}{m_b^2m_s^2m_d^2}\right)^{1/8}\right] \nn \\
&=& 4\pi v\, . \label{eq:OWLH}
\ee
So another potential trigger fails to solve the hierarchy problem above $4\pi v$.

Finally, it is worth briefly mentioning $H^\dagger H G \widetilde G$. Anything coupling to $H^\dagger H G \widetilde G$ in the SMEFT can also couple to $G \widetilde G$ with a less irrelevant coupling and the latter will dominate its cosmological dynamics. Additionally, unlike $G \widetilde G$, $H^\dagger H G \widetilde G$ is not a total derivative and gets a UV-sensitive perturbative contribution to its VEV that can be estimated following the same procedure used for the renormalization of $\bar \theta$~\cite{KHRIPLOVICH199427, Ellis:1978hq, Valenti:2022uii}. Similar arguments apply to $H^\dagger H W \widetilde W$.

\section{Conclusion}\label{sec:conclusion}
Discovering that the value of the Higgs mass is determined by a trigger operator would  change the way we think about fine-tuning problems and the cosmological evolution of our universe. In this work we circumscribe the effort of discovering or falsifying the trigger paradigm to the study of just three operators.

More precisely, we have established that SMEFT operators up to dimension 6 can only be used to solve the hierarchy problem up to $\Lambda_H \lesssim 4\pi v$. One can apply our arguments to conclude that the same is true at dimension 8. We discuss the most promising example that we found in Appendix~\ref{app:8}. Solving the hierarchy problem up to $\Lambda_H \lesssim 4\pi v$ is perfectly legitimate, but requires some amount of model building to realize the necessary cosmological dynamics, without fundamentally altering the problem that we are facing after decades of measuring only the SM. We find it more interesting to construct theories that can solve the hierarchy problem up to scales parametrically larger than those that we are already exploring at colliders. 

Those who share our point of view can focus on just three trigger operators: 1) $G \widetilde G$, 2) $H_1 H_2$, 3) $F \widetilde F$ + light vector-like leptons. $H_1 H_2$ requires the existence of new light Higgs particles within reach of HL-LHC~\cite{Arkani-Hamed:2020yna}. $F \widetilde F$ minimally requires a vector-like $SU(2)_L$ doublet below $4\pi v$, leading to interesting signatures at HL-LHC and future colliders~\cite{Beauchesne:2017ukw}. $G \widetilde G$ is the hardest to detect since it already exists in the SM. In this case the sign that a cosmological solution is realized arises in axion-like signatures of the new sector coupled to $G \widetilde G$. Since the new particles are not guaranteed to be dark matter and can span a wide range of masses (roughly of order $m_\pi f_\pi/\Lambda_H$~\cite{TitoDAgnolo:2021pjo} with $\Lambda_H$ anywhere at or below $M_{\rm Pl}$), they might be out of reach of existing or planned experiments.

Focusing solely on the three existing triggers above comes with three main caveats. First of all, triggers might exist at dimension 10 or above. However, using a high-dimensional trigger requires building a UV model that explains why it is generated and, more importantly, why it is the only coupling between us and the sector solving the hierarchy problem. Any other non-derivative coupling would jeopardize the selection of $m_h^2$ by introducing a sensitivity to larger UV scales. This is straightforward to do for $G\widetilde G$ and $F \widetilde F$, since they could be the only axion-like SM couplings of a new pseudo-scalar contributing to its potential. It is also easy to do for $H_1 H_2$ by means of a $\mathbb{Z}_2$ or $\mathbb{Z}_4$ symmetry~\cite{Arkani-Hamed:2020yna}. Overall it is so easy to construct a solution to the hierarchy problem with existing triggers because they are low-dimensional operators, which makes them natural candidates to be the leading low-energy coupling of a new particle. It becomes increasingly difficult to justify a similar construction as we increase the dimension of the operator.

A second point to keep in mind is that new BSM triggers exist. The counter-argument here is that they are typically experimentally excluded. $H_1 H_2$ is already borderline excluded~\cite{Arkani-Hamed:2020yna} and is just adding to the SM a new $SU(2)_L$ doublet. In this work we do not aim to prove that no other viable BSM trigger exists. Rather, we hope that the readers will find a new one, stimulated by our skepticism.

A third issue to consider is that we never computed $O(1)$ factors in our estimates of multi-loop diagrams. Therefore, it is possible that the maximal cutoff be larger than $4\pi v$ for some operators with $d\leq 8$. However, this is quite unlikely to alter our estimates, because the coefficients that we did not compute enter the computation of the cutoff with small fractional powers or have to be highly hierarchical between almost identical diagrams to invalidate our estimates. These points are illustrated in more detail in Section~\ref{sec:SMEFT}.

It is also important to mention that many cosmological solutions to the hierarchy problem do not require a trigger~\cite{Benevedes:2025qwt, Dvali:2003br,Dvali:2004tma,Geller:2018xvz,Cheung:2018xnu,Giudice:2021viw, Matsedonskyi:2023tca, Chattopadhyay:2024rha, Jung:2021cps, Agrawal:1997gf,Hall:2014dfa,DAmico:2019hih,Arkani-Hamed:2004ymt}. Most of those which do not require a trigger fall in what~\cite{TitoDAgnolo:2021pjo} called {\it statistical}~\cite{Dvali:2003br,Dvali:2004tma,Geller:2018xvz,Cheung:2018xnu,Giudice:2021viw, Matsedonskyi:2023tca, Chattopadhyay:2024rha, Jung:2021cps} and {\it anthropic}~\cite{Agrawal:1997gf,Hall:2014dfa,DAmico:2019hih,Arkani-Hamed:2004ymt} solutions to the problem. The first category aims at populating the Multiverse predominantly with patches with small $\hv \simeq v$. All existing solutions require eternal inflation and have a problem of measure.  Anthropic solutions require accepting a given definition of what is an observer, but once this is done they do not have a problem of measure. In summary, excluding the three trigger operators above does not entirely rule out the possibility that $\hv$ is explained cosmologically. However it is still relevant to a large class of solutions to the problem~\cite{Graham:2015cka, Espinosa:2015eda, Arvanitaki:2016xds, Geller:2018xvz,  Cheung:2018xnu, Giudice:2019iwl, Arkani-Hamed:2020yna, Strumia:2020bdy, Csaki:2020zqz, TitoDAgnolo:2021nhd, TitoDAgnolo:2021pjo, Khoury:2021zao, Chatrchyan:2022pcb, Trifinopoulos:2022tfx, Csaki:2022zbc, Matsedonskyi:2023tca, Hook:2023yzd, Chattopadhyay:2024rha, Csaki:2024ywk, Chatrchyan:2022dpy}. 

In spite of the caveats that we have just discussed, exploring the signatures of the three existing triggers is a worthwhile effort. The collider and axion experimental programs still have much to say about the hierarchy problem. 

\section*{Acknowledgments}

We would like to thank Nima Arkani-Hamed, Roberto Contino, Angelo Esposito, Gudrun Hiller and Marcello Romano for very useful discussions.
PS and RTD acknowledge ANR grant ANR-23-CE31-0024 EUHiggs for partial support. This research was supported in part by grant NSF PHY-2309135 to the Kavli Institute for Theoretical Physics (KITP). PS acknowledges financial support from the Spanish Ministry of Science and Innovation (MICINN) through the Spanish State Research Agency, under Severo Ochoa Centres of Excellence Programme 2025-2029 (CEX2024001442-S). This work is also part of the R\&D\&i project PID2023-146686NB-C31, funded by MICIU/AEI/10.13039/501100011033/ and by ERDF/EU.

\appendix

\section{$\Lambda_{\rm QCD}$ sensitivity to the Higgs VEV}\label{app:LQCDvsHiggsVEV}

In an $SU(N_c)$ gauge theory, with $N_f$ Dirac fermions in the fundamental representation, we have the one-loop $\beta$-function 
\begin{equation}
    \frac{dg_{(N_f)}}{d\log \mu}\equiv\beta(g_{(N_f)})=-\frac{b_0^{(N_f)}}{16\pi^2}g_{(N_f)}^3\, ,\quad b_0^{(N_f)}=\frac{11}{3}N_c-\frac{2}{3}N_f\, .
\end{equation}
The associated RG-invariant scale
\begin{equation}
    \Lambda_{(N_f)}=\mu \exp\left[-\frac{8\pi^2}{b_0^{(N_f)}g_{(N_f)}^2(\mu)} \right]\, ,
\end{equation}
is generated by dimensional transmutation.
When a heavy quark $Q$ of mass $m_Q$ is integrated out at the threshold $\mu=m_Q$, we require the one-loop threshold matching given by the continuity of the couplings of the two theories at that scale $g_{(N_f+1)}(m_Q)=g_{(N_f)}(m_Q)$. Equivalently, the $\Lambda$ scales above and below the threshold obey
\begin{equation}
    \Lambda_{(N_f)}=\left[\Lambda_{(N_f+1)}\right]^{\frac{b_0^{(N_f+1)}}{b_0^{(N_f)}}}\big[m_Q\big]^{1-\frac{b_0^{(N_f+1)}}{b_0^{(N_f)}}}\, .
\end{equation}
Since $b_0^{(N_f+1)}<b_0^{(N_f)}$, the exponent of $m_Q=y_Q v$, and hence of $v$ for a fixed Yukawa coupling $y_Q$, is positive.

Specializing to QCD ($N_c=3$) and integrating out, in sequence, the top, bottom, and charm quarks at thresholds $\mu\simeq m_t,\, m_b,\, m_c$, we obtain
\begin{equation}
    \log \Lambda_{(3)}=\frac{b_0^{(6)}}{b_0^{(3)}}\log\Lambda_{(6)}+\frac{b_0^{(5)}-b_0^{(6)}}{b_0^{(3)}}\log m_t+\frac{b_0^{(4)}-b_0^{(5)}}{b_0^{(3)}}\log m_b+\frac{b_0^{(3)}-b_0^{(4)}}{b_0^{(3)}}\log m_c\, ,
\end{equation}
with $b_0^{(6)}=7$, $b_0^{(5)}=23/3$, $b_0^{(4)}=25/3$, $b_0^{(3)}=9$. This yields the standard one-loop relation
\begin{equation}
    \Lambda_{(3)}=\Lambda_{(6)}^{\frac{7}{9}}m_t^{\frac{2}{27}}m_b^{\frac{2}{27}}m_c^{\frac{2}{27}}\, . 
\end{equation}
Treating Yukawas as fixed and using $m_Q\propto v$, the sensitivity of the three-flavor scale to the Higgs VEV is
\begin{equation}
    \frac{d\log{\Lambda}_{(3)}}{d\log v}=\frac{2}{27}+\frac{2}{27}+\frac{2}{27}=\frac{2}{9}\simeq 0.222\, . \label{eq:LQCD}
\end{equation}
Here we used that $\Lambda_{(6)}$ is $v$-independent at one loop (the running above $m_t$ is independent of $v$ at one loop). From Eq.~\eqref{eq:LQCD} and the VEV of $G\widetilde G$ discussed in the main text and in~\cite{Arkani-Hamed:2020yna},
\begin{equation}
    \langle G \widetilde{G}\rangle \propto \bar{\theta}\, m_{\pi}^2f_{\pi}^2\sim \bar{\theta}\,(m_u+m_d)|\langle \bar{q}q\rangle| \sim\bar{\theta}\,v\,\Lambda_{\rm QCD}^3\, 
\end{equation}
we obtain 
\begin{equation}
    \frac{d \log \langle G\widetilde{G}\rangle}{d\log v}\simeq 1+3\frac{d\log\Lambda_{(3)}}{d\log v}=\frac{5}{3}\simeq 1.67\, , 
\end{equation}
up to subleading corrections from higher-order chiral effects, higher-loop running/matching, and mild scheme dependence at two loops and above.

\section{Selected examples}\label{app:examples}
In this appendix we give more details on the Weinberg gluonic operator discussed in Section~\ref{sec:noT} and another dimension 6 QCD condensate.

In the following, we write the Higgs doublet as $H=\left(\chi^+,[v+h+i\chi^0]/\sqrt{2}\right)^T$
and, for convenience, we work in unitary gauge.

\begin{itemize}[label=\textcolor{linkcolor}{\textbullet}]

\item\textcolor{linkcolor}{$\mO_{QGH}=(Q^\dag\bar\sigma^{\mu\nu}T^Ad^{c\dag}) H G_{\mu\nu}^A$}\\
We first rewrite the operator more explicitly as
   \begin{align}
\mO_{QGH}=(Q^\dagger\bar\sigma^{\mu\nu}\bm{G}_{\mu\nu}d^{c\dag}) H=\frac{h+v}{\sqrt2}(d^\dagger\bar\sigma^{\mu\nu}\bm{G}_{\mu\nu}d^{c\dag})\,.
\end{align}
Then we note that $(d^\dagger\bar{\sigma}^{\mu\nu}\bm{G}_{\mu\nu}d^{c\dag})$ can condense from QCD non-perturbative dynamics (see e.g.~\cite{Shifman:1978bx,Shifman:1978by}),
\begin{equation}
    \langle Q\sigma^{\mu\nu}\bm{G}_{\mu\nu}d^c \rangle=m_0^2\langle Q d^c\rangle\,,~~m_0^2({\rm{GeV}^2})=0.8\pm0.2\, . \label{eq:m0}
\end{equation}
Finally we estimate the perturbative contribution to the VEV from diagrams with the same topology as those in Fig.~\ref{fig:O2}. Putting it all together, we obtain
\begin{equation}
    \langle\mO_{QGH}\rangle= m_0^2\,\Lambda_{\rm{QCD}}f_\pi^2+g_sy_d\left(c_{QGH}^{(1)}\frac{\Lh^6}{\Loop^3}+c_{QGH}^{(2)}v^2\frac{\Lh^4}{\Loop^2}+c_{QGH}^{(3)}v^4\frac{\Lh^2}{\Loop^2}\right)+\ldots\, ,
\end{equation}
where $c_{QGH}^{(i)}$ are $O(1)$ numbers. For the cutoff this means
\begin{equation}
\Lh\lesssim 4\pi v\,{\rm{max}}\left[1,\left( \frac{m_0^2\Lambda_{\rm{QCD}} f_\pi^2}{y_dg_s v^5}\right)^{1/6}\right]=4\pi v\,.
\end{equation}
\item \textcolor{linkcolor}{$\mathcal{O}_{\mathcal{W}}=f^{abc}\varepsilon^{\alpha\beta\gamma\delta}G^a_{\mu\alpha}(x)G^b_{\beta\gamma}(x)G_\delta^{\mu,c}(x)$}

\begin{figure}[t!]
    \centering

\tikzset{every picture/.style={line width=1pt}} 

\begin{tikzpicture}[x=1pt,y=1pt,yscale=-1,xscale=1]

\draw    (350.76,7.99) -- (348.56,8.08) ;
\draw [shift={(345.56,8.2)}, rotate = 357.61] [fill={rgb, 255:red, 0; green, 0; blue, 0 }  ][line width=0.08]  [draw opacity=0] (6.25,-3) -- (0,0) -- (6.25,3) -- cycle    ;
\draw   (324.71,33) .. controls (324.71,19.19) and (335.9,8) .. (349.71,8) .. controls (363.52,8) and (374.71,19.19) .. (374.71,33) .. controls (374.71,46.81) and (363.52,58) .. (349.71,58) .. controls (335.9,58) and (324.71,46.81) .. (324.71,33) -- cycle ;
\draw    (330.91,16.54) .. controls (333.26,16.4) and (334.51,17.51) .. (334.65,19.86) .. controls (334.79,22.21) and (336.04,23.32) .. (338.39,23.18) .. controls (340.74,23.04) and (341.99,24.15) .. (342.13,26.5) .. controls (342.26,28.85) and (343.51,29.96) .. (345.86,29.82) .. controls (348.21,29.68) and (349.46,30.79) .. (349.6,33.14) .. controls (349.74,35.49) and (350.99,36.6) .. (353.34,36.46) .. controls (355.69,36.32) and (356.94,37.43) .. (357.08,39.78) .. controls (357.22,42.13) and (358.47,43.24) .. (360.82,43.1) .. controls (363.17,42.96) and (364.42,44.07) .. (364.55,46.42) .. controls (364.69,48.77) and (365.94,49.88) .. (368.29,49.74) -- (368.51,49.94) -- (368.51,49.94) ;
\draw    (331.31,50.35) .. controls (331.35,48) and (332.55,46.84) .. (334.91,46.88) .. controls (337.26,46.92) and (338.46,45.76) .. (338.51,43.41) .. controls (338.56,41.06) and (339.76,39.9) .. (342.11,39.95) .. controls (344.46,39.99) and (345.66,38.83) .. (345.71,36.48) -- (347.31,34.94) -- (347.31,34.94) ;
\draw    (351.31,31.15) .. controls (351.35,28.8) and (352.55,27.64) .. (354.91,27.68) .. controls (357.26,27.72) and (358.46,26.56) .. (358.51,24.21) .. controls (358.56,21.86) and (359.76,20.7) .. (362.11,20.75) .. controls (364.46,20.79) and (365.66,19.63) .. (365.71,17.28) -- (367.31,15.74) -- (367.31,15.74) ;
\draw    (324.71,33) ;
\draw [shift={(324.76,35.86)}, rotate = 269] [fill={rgb, 255:red, 0; green, 0; blue, 0 }  ][line width=0.08]  [draw opacity=0] (6.25,-3) -- (0,0) -- (6.25,3) -- cycle    ;
\draw    (351.31,57.94) ;
\draw [shift={(353.16,57.86)}, rotate = 177.74] [fill={rgb, 255:red, 0; green, 0; blue, 0 }  ][line width=0.08]  [draw opacity=0] (6.25,-3) -- (0,0) -- (6.25,3) -- cycle    ;
\draw    (374.91,31.34) ;
\draw [shift={(374.91,30.14)}, rotate = 90] [fill={rgb, 255:red, 0; green, 0; blue, 0 }  ][line width=0.08]  [draw opacity=0] (6.25,-3) -- (0,0) -- (6.25,3) -- cycle    ;

\draw (340.89,15.24) node [anchor=north west][inner sep=0.75pt]   {$W$};
\draw (349.71,-7) node [anchor=north] {$s$};   
\draw (320,33) node [anchor=east] {$u$};       
\draw (379,33) node [anchor=west] {$c$};       
\draw (349.71,75) node [anchor=south] {$b$};   

\end{tikzpicture}

    \caption{Three-loop diagram contributing to the Weinberg gluonic operator $\mathcal{O}_{\mathcal{W}}$. }
    \label{fig:placeholder}
\end{figure}
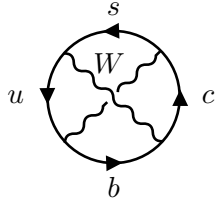

Computing the VEV of the operator $\mathcal{O}_{\mathcal W}$ using Standard Model vertices is equivalent to determining how $\mathcal{O}_{\mathcal W}$ is generated within the Standard Model. As a result of integration over quark and $W$-boson modes, one can obtain the following CP-odd effective action~\cite{Pospelov:1994uf}, 
\begin{equation}
    S^{\rm{eff}}_{\cancel{\rm{CP}}}=\int\,d^4x\left(C_1g_s^2G\widetilde G+C_2g_s^3\mathcal{O}_{\mathcal{W}}(x)+\dots\right)\,,
\end{equation}
where $C_1$ is computed in~\cite{Ellis:1978hq,KHRIPLOVICH199427} and $C_2$ is generated at three loops~\cite{Pospelov:1994uf},
\begin{equation}\label{eq:C2}
    C_2=\frac{
    \mathcal JG_F^2}{1536\pi^6}\frac{m_b^2m_c^2m_s^2}{m_W^4}\log\left(\frac{m_b^2}{m_s^2}\right)\log\left(\frac{m_W^2}{m_b^2}\right)\mathcal{I}\left(\frac{m_t^2}{m_W^2}\right)\,,
\end{equation}
where $\mathcal{J}={\rm Im }\left(V_{us}V_{cb}V_{ub}^*V_{cs}^*\right)$, $V$ is the CKM matrix and we introduced the $O(1)$ loop function
\begin{align}
    \mathcal{I}(x)&\equiv\frac{x^2}{(x-1)^4}\log x\left(3+\frac{12}{x-1}+\frac{10}{(x-1)^2}\right)\nn\\
    &-\frac{x}{(x-1)^2}\left(3+\frac{13}{x-1}+\frac{5}{(x-1)^2}+\frac{10}{(x-1)^3}\right)\,.
\end{align}
A representative diagram for $C_2$ is shown in Fig.~\ref{fig:placeholder}.
Some of the properties of $C_2$ in Eq.~\eqref{eq:C2} follow from simple spurionic arguments. Indeed, the only \textit{perturbative} flavor-invariant source of CP-violation in the SM is $\mathcal{J}$. The $\mathcal{J}$ invariant contains four CKM matrix elements, forcing the exchange of two $W$-bosons. Consequently this contribution contains four powers of the weak coupling $g$ and has (at least) a three-loop suppression. Moreover, from the unitarity of the CKM matrix we obtain the GIM suppression factor $\frac{m_b^2m_c^2m_s^2}{m_W^4}$. In addition to the perturbative contribution that we have just discussed, the Weinberg gluonic operator condenses. We compute the contribution to its VEV from QCD in Appendix~\ref{app:Weinberg}. Overall we find the results in Eqs.~\eqref{eq:OW} and~\eqref{eq:OWLH}. 
It is also worth checking explicitly that the Weinberg operator, when closed with itself, produces a non-zero VEV. To get the Feynman rules associated with the operator $\mathcal{O}_{\mathcal{W}}$, we expand
\begin{align}
    \mathcal{O}_{\mathcal{W}}(x)&=
    f^{abc}\varepsilon^{\alpha\beta\gamma\delta}G^a_{\mu\alpha}(x)G^b_{\beta\gamma}(x)G_\delta^{\mu,c}(x)\nn\\
    &=-
    2f^{abc}\epsilon^{\alpha\beta\gamma\delta}\partial_\beta A^b_\gamma\left[\partial_\mu A_\alpha^a\partial^\mu A_\delta^c-2\partial_\alpha A^a_\mu\partial^\mu A_\delta^c\right]+O(g_sA^4)\,,
\end{align}
where we omitted the contributions from higher orders in the strong coupling. We thus get the Feynman rule (for outgoing momenta)
\begin{align*}
&\begin{adjustbox}{max width=0.94\textwidth}
\raisebox{-11mm}{     
\tikzset{every picture/.style={line width=1.2pt}} 
\begin{tikzpicture}[x=0.75pt,y=0.75pt,yscale=-1,xscale=1]
\draw    (150.51,133.94) .. controls (151.26,131.71) and (152.75,130.97) .. (154.98,131.72) .. controls (157.21,132.47) and (158.71,131.72) .. (159.46,129.49) .. controls (160.21,127.26) and (161.71,126.51) .. (163.94,127.26) .. controls (166.17,128.01) and (167.67,127.27) .. (168.42,125.04) .. controls (169.17,122.81) and (170.66,122.06) .. (172.89,122.81) .. controls (175.12,123.56) and (176.62,122.82) .. (177.37,120.59) .. controls (178.12,118.36) and (179.62,117.61) .. (181.85,118.36) .. controls (184.08,119.11) and (185.57,118.37) .. (186.32,116.14) .. controls (187.07,113.91) and (188.57,113.16) .. (190.8,113.91) .. controls (193.03,114.66) and (194.53,113.92) .. (195.28,111.69) -- (199.33,109.67) -- (199.33,109.67) ;
\draw    (150.5,133.42) .. controls (152.77,132.79) and (154.22,133.61) .. (154.85,135.88) .. controls (155.48,138.15) and (156.94,138.97) .. (159.21,138.34) .. controls (161.48,137.71) and (162.93,138.53) .. (163.56,140.8) .. controls (164.19,143.07) and (165.64,143.89) .. (167.91,143.26) .. controls (170.18,142.63) and (171.64,143.45) .. (172.27,145.72) .. controls (172.9,147.99) and (174.35,148.81) .. (176.62,148.18) .. controls (178.89,147.55) and (180.34,148.36) .. (180.97,150.63) .. controls (181.6,152.9) and (183.06,153.72) .. (185.33,153.09) .. controls (187.6,152.46) and (189.05,153.28) .. (189.68,155.55) .. controls (190.31,157.82) and (191.76,158.64) .. (194.03,158.01) .. controls (196.3,157.38) and (197.76,158.2) .. (198.39,160.47) -- (199.33,161.01) -- (199.33,161.01) ;
\draw    (101,133.34) .. controls (102.69,131.69) and (104.35,131.71) .. (106,133.4) .. controls (107.65,135.09) and (109.31,135.11) .. (111,133.46) .. controls (112.69,131.81) and (114.35,131.83) .. (116,133.52) .. controls (117.65,135.21) and (119.31,135.23) .. (121,133.58) .. controls (122.69,131.93) and (124.35,131.95) .. (126,133.64) .. controls (127.65,135.33) and (129.31,135.35) .. (131,133.7) .. controls (132.69,132.05) and (134.35,132.07) .. (136,133.76) .. controls (137.65,135.45) and (139.31,135.47) .. (141,133.83) .. controls (142.69,132.18) and (144.35,132.2) .. (146,133.89) -- (150.51,133.94) -- (150.51,133.94) ;
\draw[fill=black]
    ([xshift=-2.45pt, yshift=-2.45pt]150.5,133.42)
    rectangle
    ([xshift= 2.45pt, yshift= 2.45pt]150.5,133.42);
\draw (58.08,111.88) node [anchor=north west][inner sep=0.75pt]  [font=\footnotesize]  {$\{p_{1} ,\mu _{1} ,a_{1}\}$};
\draw (202.26,162.5) node [anchor=north west][inner sep=0.75pt]  [font=\footnotesize]  {$\{p_{2} ,\mu _{2} ,a_{2}\}$};
\draw (201.59,92.26) node [anchor=north west][inner sep=0.75pt]  [font=\footnotesize]  {$\{p_{3} ,\mu _{3} ,a_{3}\}$};
\draw (136,109.73) node [anchor=north west][inner sep=0.75pt]  [font=\small]  {$\mathcal{O}_{\mathcal{W}}$};
\end{tikzpicture}
}
\end{adjustbox}\nn\\
&= 
-4if_{a_1a_2a_3}\left\{\epsilon^{\mu_1\mu_2\alpha\beta}\left[p_1^{\mu_3}{p_2}_\alpha{p_3}_\beta+{p_1}_\alpha p_2^{\mu_3} {p_3}_\beta\right]+\epsilon^{\mu_1\mu_2\mu_3\alpha}{p_1}_\alpha (p_2\cdot p_3)+\text{cyclic perm.s}\right\}\,,
\end{align*}
and one readily verifies that the sum over all possible contractions is non-vanishing.

\end{itemize}
\section{Dimension Eight}\label{app:8}

\begin{figure}[t!]
    \centering

\tikzset{every picture/.style={line width=1pt}} 

\begin{tikzpicture}[x=1pt,y=1pt,yscale=-1,xscale=1]

\draw  [fill={rgb, 255:red, 0; green, 0; blue, 0 }  ,fill opacity=1 ] (273.25,39.19) -- (278.15,39.19) -- (278.15,44.09) -- (273.25,44.09) -- cycle ;
\draw    (246.05,23.79) -- (243.85,23.88) ;
\draw [shift={(240.85,24)}, rotate = 357.61] [fill={rgb, 255:red, 0; green, 0; blue, 0 }  ][line width=0.08]  [draw opacity=0] (6.25,-3) -- (0,0) -- (6.25,3) -- cycle    ;
\draw    (241.55,59.24) ;
\draw [shift={(241.55,59.24)}, rotate = 0] [fill={rgb, 255:red, 0; green, 0; blue, 0 }  ][line width=0.08]  [draw opacity=0] (6.25,-3) -- (0,0) -- (6.25,3) -- cycle    ;
\draw    (305.35,23.79) -- (307.1,23.81) ;
\draw [shift={(310.1,23.86)}, rotate = 180.87] [fill={rgb, 255:red, 0; green, 0; blue, 0 }  ][line width=0.08]  [draw opacity=0] (6.25,-3) -- (0,0) -- (6.25,3) -- cycle    ;
\draw    (214.02,37.34) -- (219.3,43.05) ;
\draw    (219.3,37.34) -- (214.02,43.05) ;

\draw   (216.4,41.64) .. controls (216.4,31.78) and (229.67,23.79) .. (246.05,23.79) .. controls (262.43,23.79) and (275.7,31.78) .. (275.7,41.64) .. controls (275.7,51.49) and (262.43,59.49) .. (246.05,59.49) .. controls (229.67,59.49) and (216.4,51.49) .. (216.4,41.64) -- cycle ;
\draw    (310.1,59.2) ;
\draw [shift={(310.1,59.2)}, rotate = 180] [fill={rgb, 255:red, 0; green, 0; blue, 0 }  ][line width=0.08]  [draw opacity=0] (6.25,-3) -- (0,0) -- (6.25,3) -- cycle    ;
\draw    (332.49,38.55) -- (337.77,44.26) ;
\draw    (337.77,38.55) -- (332.49,44.26) ;

\draw   (275.7,41.64) .. controls (275.7,31.78) and (288.97,23.79) .. (305.35,23.79) .. controls (321.73,23.79) and (335,31.78) .. (335,41.64) .. controls (335,51.49) and (321.73,59.49) .. (305.35,59.49) .. controls (288.97,59.49) and (275.7,51.49) .. (275.7,41.64) -- cycle ;
\draw  [fill={rgb, 255:red, 0; green, 0; blue, 0 }  ,fill opacity=1 ] (424.75,39.19) -- (429.65,39.19) -- (429.65,44.09) -- (424.75,44.09) -- cycle ;
\draw    (417.73,28.76) -- (416.93,28.33) ;
\draw [shift={(414.27,26.94)}, rotate = 27.76] [fill={rgb, 255:red, 0; green, 0; blue, 0 }  ][line width=0.08]  [draw opacity=0] (6.25,-3) -- (0,0) -- (6.25,3) -- cycle    ;
\draw    (434.99,29.34) -- (436.38,28.57) ;
\draw [shift={(439,27.12)}, rotate = 151.03] [fill={rgb, 255:red, 0; green, 0; blue, 0 }  ][line width=0.08]  [draw opacity=0] (6.25,-3) -- (0,0) -- (6.25,3) -- cycle    ;
\draw   (367.9,41.64) .. controls (367.9,31.78) and (381.17,23.79) .. (397.55,23.79) .. controls (413.93,23.79) and (427.2,31.78) .. (427.2,41.64) .. controls (427.2,51.49) and (413.93,59.49) .. (397.55,59.49) .. controls (381.17,59.49) and (367.9,51.49) .. (367.9,41.64) -- cycle ;
\draw    (460.63,59.2) ;
\draw [shift={(460.63,59.2)}, rotate = 180] [fill={rgb, 255:red, 0; green, 0; blue, 0 }  ][line width=0.08]  [draw opacity=0] (6.25,-3) -- (0,0) -- (6.25,3) -- cycle    ;
\draw   (427.2,41.64) .. controls (427.2,31.78) and (440.47,23.79) .. (456.85,23.79) .. controls (473.23,23.79) and (486.5,31.78) .. (486.5,41.64) .. controls (486.5,51.49) and (473.23,59.49) .. (456.85,59.49) .. controls (440.47,59.49) and (427.2,51.49) .. (427.2,41.64) -- cycle ;

\draw  [fill={rgb, 255:red, 0; green, 0; blue, 0 }  ,fill opacity=1 ] (395.9,23.79) .. controls (395.9,22.87) and (396.64,22.14) .. (397.55,22.14) .. controls (398.46,22.14) and (399.2,22.87) .. (399.2,23.79) .. controls (399.2,24.7) and (398.46,25.44) .. (397.55,25.44) .. controls (396.64,25.44) and (395.9,24.7) .. (395.9,23.79) -- cycle ;
\draw  [fill={rgb, 255:red, 0; green, 0; blue, 0 }  ,fill opacity=1 ] (455.2,23.79) .. controls (455.2,22.87) and (455.94,22.14) .. (456.85,22.14) .. controls (457.76,22.14) and (458.5,22.87) .. (458.5,23.79) .. controls (458.5,24.7) and (457.76,25.44) .. (456.85,25.44) .. controls (455.94,25.44) and (455.2,24.7) .. (455.2,23.79) -- cycle ;
\draw    (378.82,27.85) -- (380.21,27.3) ;
\draw [shift={(383,26.21)}, rotate = 158.63] [fill={rgb, 255:red, 0; green, 0; blue, 0 }  ][line width=0.08]  [draw opacity=0] (6.25,-3) -- (0,0) -- (6.25,3) -- cycle    ;
\draw    (475.36,27.8) ;
\draw [shift={(473,26.53)}, rotate = 28.3] [fill={rgb, 255:red, 0; green, 0; blue, 0 }  ][line width=0.08]  [draw opacity=0] (6.25,-3) -- (0,0) -- (6.25,3) -- cycle    ;
\draw    (407.84,58.19) .. controls (405.8,56.88) and (405.47,55.2) .. (406.86,53.15) .. controls (408.53,51.68) and (408.71,50.04) .. (407.4,48.23) .. controls (406.8,45.82) and (407.75,44.53) .. (410.25,44.36) .. controls (412.39,45.05) and (413.91,44.36) .. (414.82,42.3) .. controls (416.23,40.33) and (417.89,40.09) .. (419.78,41.57) .. controls (421.43,43.18) and (423.1,43.17) .. (424.78,41.52) -- (427.2,41.64) ;
\draw    (397.55,59.49) ;
\draw [shift={(394.67,59.6)}, rotate = 357.64] [fill={rgb, 255:red, 0; green, 0; blue, 0 }  ][line width=0.08]  [draw opacity=0] (6.25,-3) -- (0,0) -- (6.25,3) -- cycle    ;
\draw    (416.75,55.06) ;
\draw [shift={(414.75,56.06)}, rotate = 333.43] [fill={rgb, 255:red, 0; green, 0; blue, 0 }  ][line width=0.08]  [draw opacity=0] (6.25,-3) -- (0,0) -- (6.25,3) -- cycle    ;
\draw    (256.34,58.19) .. controls (254.3,56.88) and (253.97,55.2) .. (255.36,53.15) .. controls (257.03,51.68) and (257.21,50.04) .. (255.9,48.23) .. controls (255.3,45.82) and (256.25,44.53) .. (258.75,44.36) .. controls (260.89,45.05) and (262.41,44.36) .. (263.32,42.3) .. controls (264.73,40.33) and (266.39,40.09) .. (268.28,41.57) .. controls (269.93,43.18) and (271.6,43.17) .. (273.28,41.52) -- (275.7,41.64) ;
\draw    (264.75,55.23) ;
\draw [shift={(262.75,56.23)}, rotate = 333.43] [fill={rgb, 255:red, 0; green, 0; blue, 0 }  ][line width=0.08]  [draw opacity=0] (6.25,-3) -- (0,0) -- (6.25,3) -- cycle    ;
\draw  [draw opacity=0][dash pattern={on 0.84pt off 2.51pt}] (396.93,23.76) .. controls (403.04,17.91) and (414.22,14) .. (427,14) .. controls (439.69,14) and (450.79,17.86) .. (456.93,23.63) -- (427,34) -- cycle ; \draw  [dash pattern={on 0.84pt off 2.51pt}] (396.93,23.76) .. controls (403.04,17.91) and (414.22,14) .. (427,14) .. controls (439.69,14) and (450.79,17.86) .. (456.93,23.63) ;  

\node at (255,18) {$u$};    
\node at (295,17.) {$d$};    
\node at (255,65) {$u^c$};  
\node at (255,40) {$g$};  
\node at (295,64.8) {$d^c$};  
\node at (426.93,20) {$h$};

\end{tikzpicture}

    \caption{Leading diagrams contributing to the VEV of $\mO_{d=8}$ in Eq.~\eqref{eq:d8}.}
    \label{fig:dim8}
\end{figure}
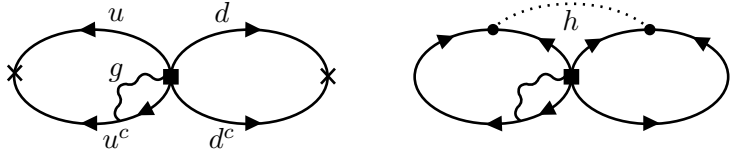
Here we discuss the dimension-8 operator from the list in~\cite{Li:2020gnx} that naively looks most promising as a trigger. It reads
\begin{equation}\label{eq:d8}
\mO_{d=8}=\epsilon^{ij}(T^A)^b_{\,\,\, a}(Q_{ri,l}u^{c,l}_t)(d_{p,b}^c\sigma^{\mu\nu}G^A_{\mu\nu}Q_{sj}^a)=\epsilon^{ij}(Q_{ri}u^c_t)(d_p^c\sigma^{\mu\nu}\bm{G}_{\mu\nu}Q_{sj})\,, 
\end{equation}
where $a$, $b$, $l$ are $SU(3)_c$ indices, $i$, $j$ are $SU(2)_L$ indices and $p,r,s,t$ are flavor indices. In the notation of~\cite{Li:2020gnx}, the operator~\eqref{eq:d8} belongs to the class $\mathcal{O}_{GQ^2u^cd^c}^{(1-12)}$. 
If we take $r=t$ and $p=s$, $\mO_{d=8}$ gets contributions to its VEV from the soft breaking of chiral symmetry by QCD and the hard breaking from Yukawa couplings (see Fig.~\ref{fig:dim8}). 
We can estimate the contribution from QCD by assuming factorization for the condensate (which is valid at leading order in the $1/N_c$ expansion), 
\begin{equation}
    \langle(Q_{ri}\bm{G}_{\mu\nu}u^c_r)(d_p^c\sigma^{\mu\nu}Q_{pj})\rangle_{\rm{QCD}}=\langle Q_{ri}\bm{G}_{\mu\nu}u^c_r\rangle_{\rm{QCD}}\langle d_p^c\sigma^{\mu\nu}Q_{pj}\rangle_{\rm{QCD}}\, .
\end{equation}
With this approximation we find
\begin{equation}    \langle\mO_{d=8}\rangle=  m_0^2\Lambda_{\rm{QCD}}^2f_\pi^4+\frac{y_u y_d g_s}{(16\pi^2)^3}\left(c_{81}v^2\Lh^6+c_{82}\frac{\Lh^8}{16\pi^2}\right)\,,
\end{equation}
where $c_{8i}$ are $O(1)$ numbers and $m_0$ is given by Eq.~\eqref{eq:m0}. We thus get
\begin{equation}
    \Lh\lesssim 4\pi v\,\text{max}\left[1,\left(\frac{\Loop^4 m_0^2\Lambda_{\rm{QCD}}^2f_\pi^4}{y_uy_dg_s}\right)^{1/8}\right]=4\pi v\,.
\end{equation}

\section{VEV of the Weinberg gluonic operator from instantons}\label{app:Weinberg}

\subsection{The Weinberg gluonic operator in the instanton background}

In this Section we estimate the VEV of the Weinberg gluonic 
\begin{equation}
    \mathcal{O}_{\mathcal{W}}(x)=f^{abc}G_{\mu\nu}^{a}G^{b\, \nu}_{\,\,\,\,\,\,\rho}\widetilde{G}^{c\, \rho\mu}\,, \quad \widetilde{G}^{a\, \mu\nu}\equiv\frac{1}{2}\varepsilon^{\mu\nu\rho\sigma}G^{a}_{\rho\sigma}\, 
\end{equation}
in QCD.
In the minimal embedding of the BPST instanton solution~\cite{Belavin:1975fg} into $SU(N)$, the instanton field strength takes the form
\begin{equation}
    G_{\mu\nu}^{a}(x)=-\frac{4}{g_s}\frac{\rho^2}{[(x-x_0)^2+\rho^2]^2}\eta^{a}_{\mu\nu}\, , \quad \widetilde{G}_{\mu\nu}^{a}=+\, G_{\mu\nu}^{a}\, ,
\end{equation}
where $\eta_{a\mu\nu}$ are the ’t Hooft symbols obeying~\cite{Shifman:2012zz}
\begin{equation}
    \eta_{a\mu\nu}\eta_{b\mu\lambda}=\delta_{ab}\delta_{\nu\lambda}+\epsilon_{abc}\eta_{c\nu\lambda}\, , \quad \eta_{a\mu\nu}\eta_{b\mu\nu}=4\delta_{ab}\, .
\end{equation}
In the embedded $SU(2)$ this implies $\epsilon^{abc}\eta_{a\mu\nu}\eta_{b\nu \rho}\eta_{c\rho\mu}=24$, and using self-duality, the Weinberg gluonic operator in the instanton background evaluates to~\cite{Bedi:2022qrd}
\begin{equation}
    \mathcal{O}_W(x)=f^{abc}G_{\mu\nu}^{a}G^{b\, \nu}_{\,\,\,\,\,\rho}G^{c\, \rho\mu}=\frac{1}{g_s^3}\frac{-1536\rho^6}{\left[(x-x_0)^2+\rho^2\right]^6}\, .
\end{equation}

\subsection{$\theta$-induced VEV for the Weinberg gluonic operator}

We wish to determine the VEV of the Weinberg operator induced by CP violation from $\bar{\theta}$. Following the standard source method, we introduce constant classical sources for the two CP-odd operators, 
\begin{equation}
    Q(x)=\frac{\alpha_s}{8\pi}G^{a}_{\mu\nu}\widetilde{G}^{a\, \mu\nu}\, , \quad  \mathcal{O}_{\mathcal{W}}(x)=f^{abc}G_{\mu\nu}^{a}G^{b\, \nu}_{\,\,\,\,\,\rho}\widetilde{G}^{c\, \rho\mu}\, ,
\end{equation}
and deform the action as
\begin{equation}
    S[\Phi;\theta,\lambda]=S_0[\Phi]+\int d^4x~\big[\bar{\theta}\, Q(x)+\lambda\, \mathcal{O}_{\mathcal{W}}(x) \big]\, .
\end{equation}
We define the generating functionals
\begin{equation}
    Z[\bar\theta,\lambda]=\int\mathcal{D}\Phi~e^{iS[\Phi;\bar\theta,\lambda]}\, , \quad W[\bar\theta,\lambda]=-i\log Z[\bar\theta,\lambda]\, .
\end{equation}
Under CP, $Q(x)\mapsto -Q(x)$ and $\mathcal{O}_{\mathcal{W}}(x)\mapsto-\mathcal{O}_{\mathcal{W}}(x)$, while the action and the path-integral measure are invariant. So we can make a field redefinition equal to a CP transformation and obtain 
\begin{equation}
    Z(\bar\theta,\lambda)=\int\mathcal{D}\Phi~e^{iS_0[\Phi]+i\bar\theta\int Q+i\lambda\int\mathcal{O}_{\mathcal{W}}}=\int\mathcal{D}\Phi~e^{iS_0[\Phi]-i\bar\theta\int Q-i\lambda\int\mathcal{O}_{\mathcal{W}}}=Z(-\bar\theta,-\lambda)\, .
\end{equation}
Hence $W(\bar\theta,\lambda)=-i\log Z(\bar\theta,\lambda)$ satisfies $W(\bar\theta,\lambda)=W(-\bar\theta,-\lambda)$. If we define an energy density $\mathcal{E}$ by factoring out of $W$ the volume of spacetime $\mathcal{V}_4$,
\be
\mathcal{E}(\bar\theta,\lambda)\equiv -W(\bar\theta,\lambda)/\mathcal{V}_4\, ,
\ee
$W(\theta, \lambda)=W(-\theta, -\lambda)$ implies that the leading terms for small $\theta$ and $\lambda$ in the energy density are
\be
\mathcal{E}(\bar\theta,\lambda)=\mathcal{E}_0+\frac{1}{2}\chi_t\bar\theta^2-\bar\theta\lambda\chi_{QW}+\frac{1}{2}C_{\mathcal{W}\mathcal{W}}\lambda^2+\ldots\, ,
\ee
where $\chi_t=\partial^2\mathcal{E}/\partial\bar\theta^2 |_0$ is the usual topological susceptibility and $C_{\mathcal{W}\mathcal{W}}=\partial^2\mathcal{E}/\partial\lambda^2 |_0$.
At this point we can show that the VEV of $\mO_{\mathcal{W}}$ is proportional to $\chi_{QW}$, since, for constant (space-time independent) sources,
\be
\frac{\partial W}{\partial\lambda}=\int d^4y~\langle \mathcal{O}_{\mathcal{W}}(y)\rangle\, \quad \Rightarrow \quad \langle\mathcal{O}_{\mathcal{W}}\rangle_{\bar\theta,\lambda}=-\frac{\partial\mathcal{E}}{\partial\lambda} = \bar\theta\chi_{QW}+O(\bar\theta^3)\, .
\ee
In the second Equation we neglected the terms proportional to $\lambda^2$ following the same logic used in Section~\ref{sec:existing} where $\lambda(x)=\xi \phi(x)$ is a probe background field with a parametrically weak coupling $\xi$. The CP-even mixed susceptibility of two CP-odd operators~\cite{Bedi:2022qrd} is given by
\begin{equation}
    \chi_{QW}=\left.\frac{\partial ^2\mathcal{E}}{\partial\bar\theta\partial \lambda}\right|_{\bar\theta=\lambda=0}=i \int d^4x~\big\langle \textrm{T}\,\mathcal{O}_{\mathcal{W}}(0)Q(x) \big\rangle_{\text{conn.}}\, .
\end{equation}
We evaluate $\chi_{QW}$ nonperturbatively with the usual instanton gas saddle point approximation of the path integral. In the semiclassical one-instanton approximation (with the standard one-loop instanton measure and including light-quark zero modes), $\chi_{QW}$ reduces to a $\rho$-integral weighted by the instanton density $n(\rho)$ and the operator insertions~\cite{Bedi:2022qrd}. Including both instanton and anti-instanton sectors (which contribute equally), in the presence of $N_f$ light quarks, one obtains~\cite{Bedi:2022qrd}
\begin{equation}
    \chi_{QW}=2\frac{384\pi^2}{5}\int_0^{\infty}\frac{d\rho}{\rho^7}~C(N,N_f)\left(\frac{8\pi^2}{g^2(1/\rho)} \right)^{2N}e^{-\frac{8\pi^2}{g^2(1/\rho)}}\prod_{f=1}^{N_f}(m_f\rho)\, ,
\end{equation}
where $C(N,N_f)$ is the standard one-loop constant from gauge-boson, ghost and fermion (non-zero modes) determinants in the $SU(N)$ instanton measure~\cite{tHooft:1976rip,tHooft:1976snw,Csaki:2019vte,Sesma:2024tcd}. This makes the $v$-dependence manifest via $m_f\propto y_f v$ and ensures $\chi_{QW}\rightarrow 0$ in the chiral limit, as required from chiral symmetry.

To expose the scaling of the $\rho$-integrand, we use the integrated RGE
\begin{equation}
    e^{-\frac{8\pi^2}{g^2(1/\rho)}}=(\rho M)^{b_0}e^{-\frac{8\pi^2}{g^2(M)}}\, , \quad b_0=\frac{11}{3}N-\frac{2}{3}N_f\, ,
\end{equation}
so the integrand behaves as $\rho^{b_0+N_f-7}$. For QCD $(N=3,N_f=2)$ this power is $14/3$, hence the integral is IR-dominated and requires an IR cutoff $\rho_{\rm max}\sim \Lambda_{\rm QCD}^{-1}$.

However it is well-known that the IR tail is not reliably captured by the dilute gas alone~\cite{Callan:1977gz,Witten:1978bc,Witten:1980sp,Vainshtein:1981wh,Coleman:1985rnk, DiVecchia:1980yfw}. A conservative, data-driven procedure is to fix 
\be
\rho_{\rm max}=c/\Lambda_{\rm QCD}
\ee
by matching the instanton expression for the topological susceptibility\footnote{We use the one-loop instanton measure with $N_f$ light quarks of masses $m_f$~\cite{tHooft:1976rip,tHooft:1976snw}
\begin{equation}
    n(\rho)=\frac{C(N,N_f)}{\rho^5}\left(\frac{8\pi^2}{g^2(1/\rho)}\right)^{2N}e^{-\frac{8\pi^2}{g^2(1/\rho)}}\prod_{f=1}^{N_f}(m_f\rho)\, .
\end{equation}} to its lattice value~\cite{Vainshtein:1981wh}:
\begin{equation}
    \chi_t=\int d^4x~\langle Q(x)Q(0)\rangle\simeq \int_0^{\rho_{\rm max}}d\rho~n(\rho)\overset{!}{=}\chi_t^{\rm latt}\, ,
\end{equation}
where $\chi_t^{\rm latt}$ at physical quark masses is given by~\cite{Borsanyi:2016ksw}
\begin{equation}
    \chi_t^{\rm latt}\simeq (75~\text{MeV})^4\simeq 3.2\times 10^{-5}~\text{GeV}^4\, .
\end{equation}
This absorbs the detailed IR physics into a single empirical number $c$ and the output will be the calibrated $\chi_{QW}$~\cite{Vainshtein:1981wh}. 
If we define the normalized size distribution
\begin{equation}
    P(\rho)=\frac{n(\rho)}{\int_0^{\rho_{\rm max}}d\rho ~n(\rho)}=\frac{n(\rho)}{\chi_t}\,, \quad \int_0^{\rho_{\rm max}}d\rho ~P(\rho)=1\, .
\end{equation}
Then the susceptibility of interest collapses to a single IR moment,
\begin{equation}
    \chi_{QW}=\frac{768\pi^2}{5}\left\langle \rho^{-2}\right\rangle_{P}\chi_t\, ,
\end{equation}
and the calibrated prediction follows by replacing $\chi_t\rightarrow \chi_t^{\rm latt}$ and using the same $\rho_{\rm max}$ that saturates $\chi_t$:
\begin{equation}
    \chi_{QW}^{(\text{cal})}=\frac{768\pi^2}{5}\left\langle\rho^{-2}\right\rangle_{P(c)}\chi_t^{\rm latt}\, .
\end{equation}
For a quick analytic estimate of $\langle\rho^{-2}\rangle_P$, with $n(\rho)\propto \rho^{a}$, $a=b_0+N_f-5$, we find
\begin{equation}
    \left\langle\rho^{-2}\right\rangle_P=\frac{a+1}{a-1}\frac{1}{\rho_{\rm max}^2}\, .
\end{equation}
For $SU(N)$ with $N_f=2$, $a=20/3$ and $\langle\rho^{-2}\rangle_P=\frac{23}{17}(\Lambda_{\rm QCD}/c)^2$ we find
\begin{equation}
    \chi_{QW}^{(\text{cal})}=\frac{17664\pi^2}{85}\frac{\Lambda_{\rm QCD}^2}{c^2}\chi_t^{\rm latt}\, .
\end{equation}
Taking $\Lambda_{\rm QCD}=330~\text{MeV}$ and, for definiteness\footnote{Ref.~\cite{Schafer:1996wv} quotes the average size $\bar{\rho}\simeq 1/3~\text{fm}$ as the standard instanton-liquid model benchmark. Lattice studies find the $SU(3)$ distribution peaked near $0.3-0.4~\text{fm}$~\cite{Hasenfratz:1998qk,Negele:1998ev,Smith:1998wt}. The choice $c=0.6$ corresponds to $\rho_{\rm max}\sim 0.35~\text{fm}=1.77~\text{GeV}^{-1}$ and $\Lambda_{\rm QCD}\sim 0.33~\text{GeV}$.}, $c=0.6$ (which corresponds to $\rho_{\rm max}\simeq 0.35~\text{fm}\simeq 1.8~\text{GeV}^{-1}$), we obtain
\begin{equation}
    \chi_{QW}^{(\text{cal})}\simeq 2\times 10^{-2}~\text{GeV}^6\, .
\end{equation}
It is often convenient to factor out the chiral piece of $\chi_t$ and define a dimensionless susceptibility. Writing $\chi_t= m_u m_d f_{\pi}^2 \widetilde{\chi}_t$ and
\begin{equation}
    \chi_{QW}=m_u m_d f_{\pi}^2 \Lambda_{\rm QCD}^2 \widetilde{\chi}_{QW}\, ,
\end{equation}
we obtain
\begin{equation}
     \widetilde{\chi}_{QW}^{(\text{cal})}=\frac{17664\pi^2}{85\,c^2}\widetilde{\chi}_t^{\rm latt} \, .
\end{equation}
This parametrization makes the mass dimensions manifest and is used in the main text.

\bibliographystyle{JHEP}
\bibliography{biblio.bib}

\end{document}